\documentclass[a4paper,fleqn,usenatbib]{mnras}

\usepackage[T1]{fontenc}
\usepackage{ae,aecompl}

\usepackage{graphicx}	
\usepackage{amsmath}	
\usepackage{amssymb}	
\usepackage{subfigure}
\usepackage{multirow}
\usepackage{xcolor}

\newcommand{\pcc}{\,{\rm cm}^{-3}}
\newcommand{\gcc}{\,{\rm g \, cm}^{-3}}
\newcommand{\pcs}{\,{\rm cm}^{-2}}
\newcommand{\um}{\, {\rm \mu m}}

\newcommand{\kel}{\, {\rm K}}

\newcommand{\pc}{\, {\rm pc}}

\newcommand{\myr}{\, {\rm Myr}}

\newcommand{\kms}{\, {\rm km \, s^{-1}}}
\newcommand{\av}{A_{\rm V}}

\title[Filament molecular lines]{Line emission from filaments in molecular clouds}

\author[Priestley et al.]{
F. D. Priestley$^1$\thanks{Email: priestleyf@cardiff.ac.uk}, D. Arzoumanian$^2$ \& A. P. Whitworth$^1$
\\
$^1$School of Physics and Astronomy, Cardiff University, Queen's Buildings, The Parade, Cardiff CF24 3AA, UK \\
$^2$Division of Science, National Astronomical Observatory of Japan, 2-21-1 Osawa, Mitaka, Tokyo 181-8588, Japan \\
}

\date{Accepted XXX. Received YYY; in original form ZZZ}

\pubyear{2023}

\begin{document}
\label{firstpage}
\pagerange{\pageref{firstpage}--\pageref{lastpage}}
\maketitle

\begin{abstract}

  Filamentary structures are often identified in column density maps of molecular clouds, and appear to be important for both low- and high-mass star formation. Theoretically, these structures are expected to form in regions where the supersonic cloud-scale turbulent velocity field converges. While this model of filament formation successfully reproduces several of their properties derived from column densities, it is unclear whether it can also reproduce their kinematic features. We use a combination of hydrodynamical, chemical and radiative transfer modelling to predict the emission properties of these dynamically-forming filaments {in the $^{13}$CO, HCN and N$_2$H$^+$ $J=1-0$ rotational lines}. The results are largely in agreement with observations; in particular, line widths are typically subsonic to transonic, even for filaments which have formed from highly supersonic inflows. If the observed filaments are formed dynamically, as our results suggest, no equilibrium analysis is possible, and simulations which presuppose the existence of a filament are likely to produce unrealistic results.

\end{abstract}

\section{Introduction}

Molecular clouds, the birthplaces of stars, are typically filamentary in nature (see \citet{pineda2022} and \citet{hacar2022} for recent reviews). The majority of prestellar cores, the objects expected to eventually form a star or stellar system, are observed to be associated with filaments \citep{konyves2015,konyves2020}, suggesting that these filaments may play a significant role in star formation. Filaments may fragment and thus subsequently form these cores \citep[e.g.][]{clarke2017}, and channel accreting material onto them \citep{peretto2013,anderson2021}. They may be formed by the interaction between gas dynamics and magnetic fields \citep{pattle2021,arzoumanian2021}. Understanding the formation of these structures is therefore necessary in order to understand star formation.

In hydrodynamical simulations of molecular clouds, filamentary structures form naturally as a result of supersonic turbulence \citep{kirk2015,federrath2016,priestley2020}, as regions where the turbulent velocity field converges tend to produce elongated overdensities which manifest as filaments {in projected maps of the column density}. This formation mechanism reproduces many of the observationally-derived properties of filaments in molecular clouds \citep{priestley2022}: the typical width of $\sim 0.1 \pc$ \citep{arzoumanian2011,arzoumanian2019}; the detailed shape of the width distribution \citep{panopoulou2017,arzoumanian2019}; the relatively flat density profiles (compared to an isothermal filament in {hydrostatic} equilibrium; \citealt{palmeirim2013,whitworth2021}); and the lack of correlation between peak surface density and filament width \citep{arzoumanian2011,arzoumanian2019}.

The agreements between theory and observation outlined above all rely on {column densities derived from} submillimetre dust continuum observations. These {column densities} rely on assumptions such as a uniform line-of-sight grain temperature, which may not be the case in reality, and could substantially distort the results \citep{howard2019,howard2021}. {Confusion between overlapping line-of-sight structures can also affect the inferred filament properties \citep[e.g.][]{juvela2012}. Dynamical properties of filaments derived from molecular line emission,} such as sub- to transonic molecular line widths \citep{arzoumanian2013,hacar2013,hacar2018,suri2019}, are not obviously consistent with a picture of filaments forming via supersonic turbulent flows. In this paper, we investigate the predicted line emission from such a model, and thereby establish {the viability of this picture} as a description of real filaments.

\section{Method}

We consider a hydrodynamical model of filament formation in a cylindrically-symmetric converging flow. A full description is given in \citet{priestley2022}; we briefly summarise the key features here. The initial state of the model is a uniform-density cylinder of isothermal gas, described by two dimensionless parameters: ${\cal G}$, the ratio of the cylinder's line density to the critical value for stability against gravitational collapse, $\mu_{\rm CRIT} = 2a_0^2/G$, where $a_0$ is the isothermal sound speed and $G$ is the gravitational constant \citep{StodolkiewiczJ1963,ostriker1964}; and ${\cal M}$, the Mach number of the radial inflow. For an initial gas density $\rho_0$, inflow velocity $v_0$, and cylinder radius $W_0$, ${\cal G} = \pi G \rho_0 W_0^2/2 a_0^2$ and ${\cal M} = v_0/a_0$. We investigate model filaments with ${\cal G} = 1.2$, which are marginally unstable to collapse, and with ${\cal M} = 1$ and $3$, corresponding to transonic and supersonic flows. Radial distance from the axis of symmetry is given by the symbol $w$, and the projected distance (i.e. for surface density profiles) by $b$.

We set the isothermal sound speed to $a_0 = 0.187 \kms$ (appropriate for molecular gas at $10 \kel$) and the cylinder radius to $W_0 = 1 \pc$, giving an initial gas density $\rho_0 = 4.2 \times 10^{-22} \gcc$, {and a hydrogen nuclei density} $n_{\rm H,0} = 180 \pcc$. The inflow velocities are then $v_0 = 0.187 \kms$ and $0.561 \kms$ for the Mach 1 and 3 models respectively. {The initial cylinder is surrounded by material with a density ten times lower and a temperature ten times higher than that within $W_0$, to ensure pressure balance across the boundary.} We simulate the evolution of these models using the {\sc phantom} smoothed-particle hydrodynamics (SPH) code \citep{price2018}, with $5 \times 10^6$ particles per model. Our SPH simulations are of a $5 \pc$ long cylinder, rather than one of infinite length, but this is a sufficient length to avoid any significant edge effects \citep[e.g.][]{clarke2015}; the inflow dominates the filament evolution, rather than gravitational contraction along the longitudinal axis.

We randomly select $10^5$ SPH particles within a $1 \pc$ radius of the simulation origin, and track their chemical evolution using the time-dependent {\sc uclchem} code \citep{holdship2017} with the UMIST12 reaction network \citep{mcelroy2013}. Elemental abundances are taken from \citet{sembach2000}, and we assume a cosmic ray ionisation rate of $1.3 \times 10^{-17} \, {\rm s^{-1}}$ and an external ultraviolet (UV) radiation field of $1.7$ Habing units \citep{habing1968}. {This UV field strength is the value typically assumed for the Solar neighbourhood \citep{jura1974,draine1978}; filaments located near sites of active star formation, such as the Orion filaments in \citet{suri2019}, may be exposed to significantly stronger external UV fields.} The column density shielding each particle from the UV field, $N_{\rm H}$, is calculated by integrating the filament density profile radially outwards from the particle position, {with a conversion factor $\av = 6 \times 10^{-22} N_{\rm H} \, {\rm cm^2 \, mag}$ from \citet{bohlin1978}}. This is certain to underestimate the true level of shielding; any sightline other than radially outward will have a higher line-of-sight column density. We therefore also consider models with no external UV field, representing the extreme limit of total shielding, although the UV field generated by cosmic ray interactions is still modelled.

Molecular line intensity position-position-velocity (PPV) cubes are created using the {\sc lime} radiative transfer code \citep{brinch2010}, with collisional data taken from the LAMDA database \citep{schoier2005} {and dust properties from \citet{ossenkopf1994} {with thin ice mantles}}. We use $10^5$ sampling points, each of which is assigned the properties of the closest post-processed SPH particle. {The spatial and velocity resolutions of the PPV cubes are $0.01 \pc$ and $0.015 \kms$ respectively.} We focus on the $J-1-0$ rotational transitions of $^{13}$CO, HCN and N$_2$H$^+$. {The $^{13}$CO line is excited in relatively low-density gas ($\sim 100 \pcc$; \citealt{clark2019}) compared to the latter two, so should trace all of the filament material to some extent. The HCN and N$_2$H$^+$ emission should preferentially come from the densest regions, as these lines have similarly high ($\gtrsim 10^4 \pcc$) critical densities \citep{shirley2015}, with the {HCN} line typically being optically thick and so potentially showing infall signatures \citep{myers1996}.} Our reaction network does not include isotopic chemistry, so we assume the $^{12}$CO/$^{13}$CO abundance ratio reflects an underlying $^{12}$C/$^{13}$C ratio of $77$ \citep{wilson1994}. HCN and N$_2$H$^+$ have hyperfine structure, which we neglect in the radiative transfer calculations {due to the significantly higher computational cost of including it}, but we {argue} in Appendix \ref{sec:hfs} that this has little impact on our {main} results.

\begin{figure*}
  \centering
  \includegraphics[width=\columnwidth]{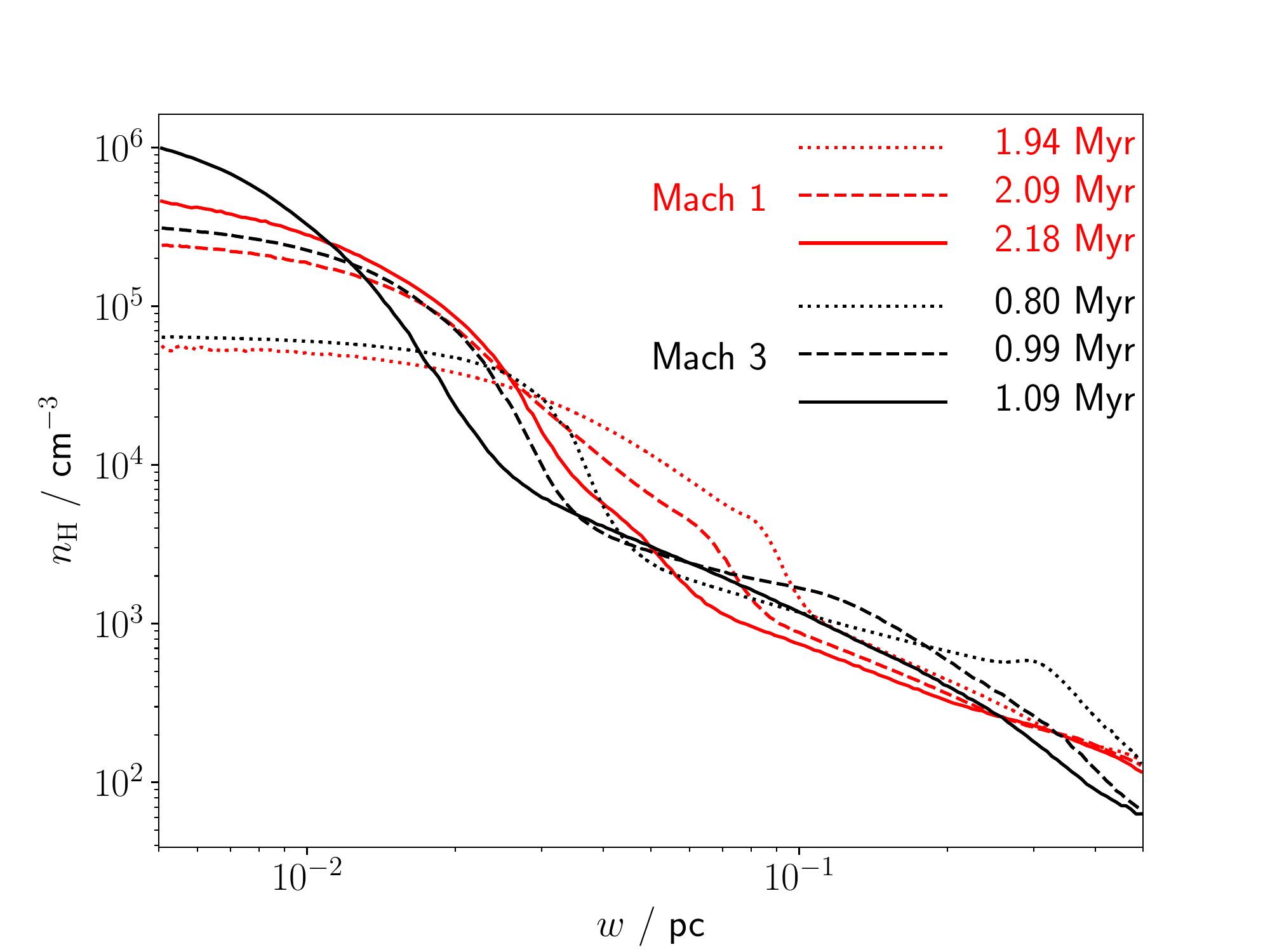}\quad
  \includegraphics[width=\columnwidth]{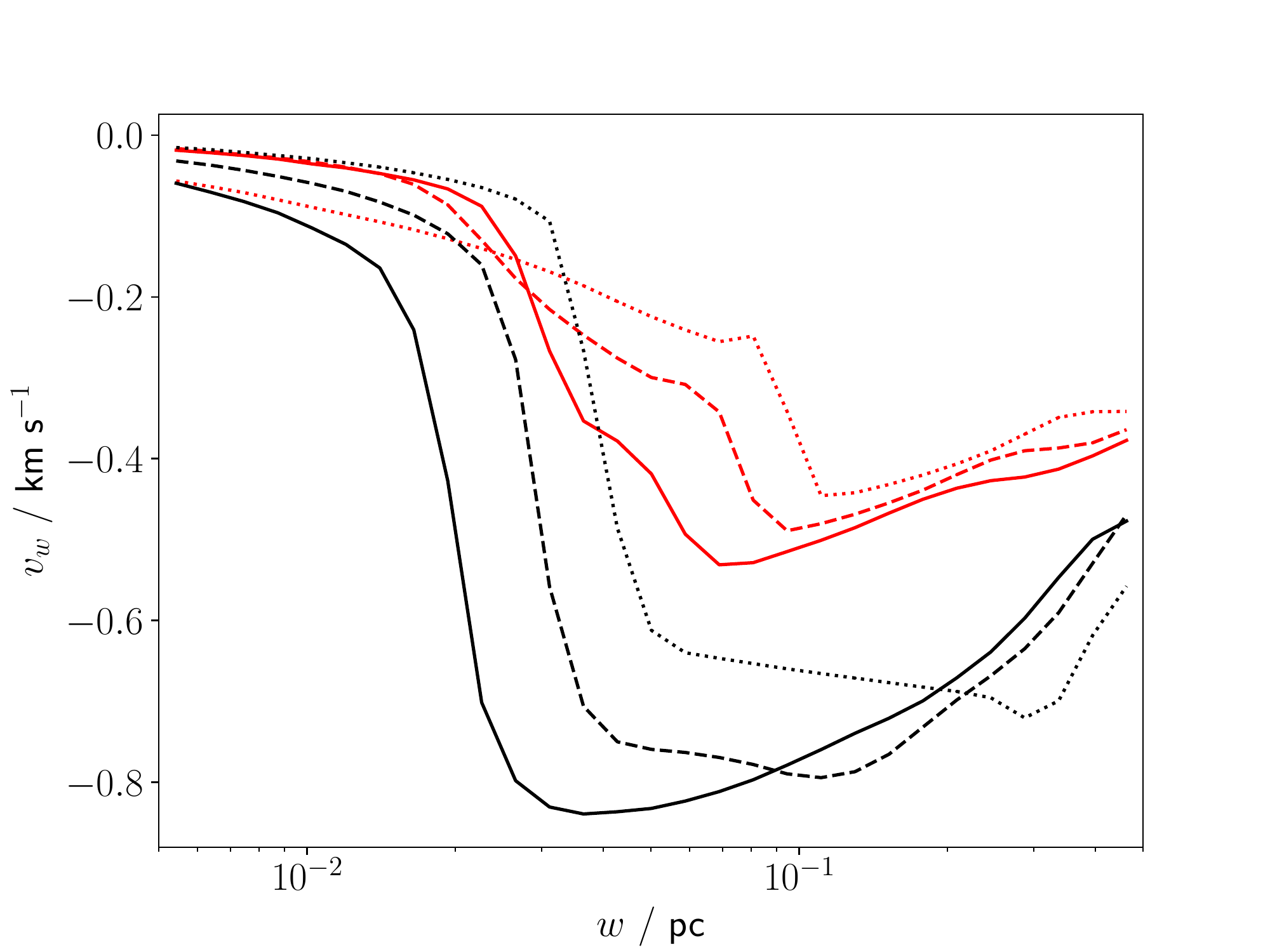}\quad
  \caption{Density (left) and radial velocity (right) profiles for the ${\cal M} = 1$ model (red) at $1.94$ (dotted line), $2.09$ (dashed line) and $2.18 \myr$ (solid line), and the ${\cal M} = 3$ model (black) at $0.80$ (dotted line), $0.99$ (dashed line) and $1.09 \myr$ (solid line). {The radial distance from the axis of symmetry is denoted by $w$.}}
  \label{fig:profile}
\end{figure*}

\begin{figure*}
  \centering
  \includegraphics[width=0.32\textwidth]{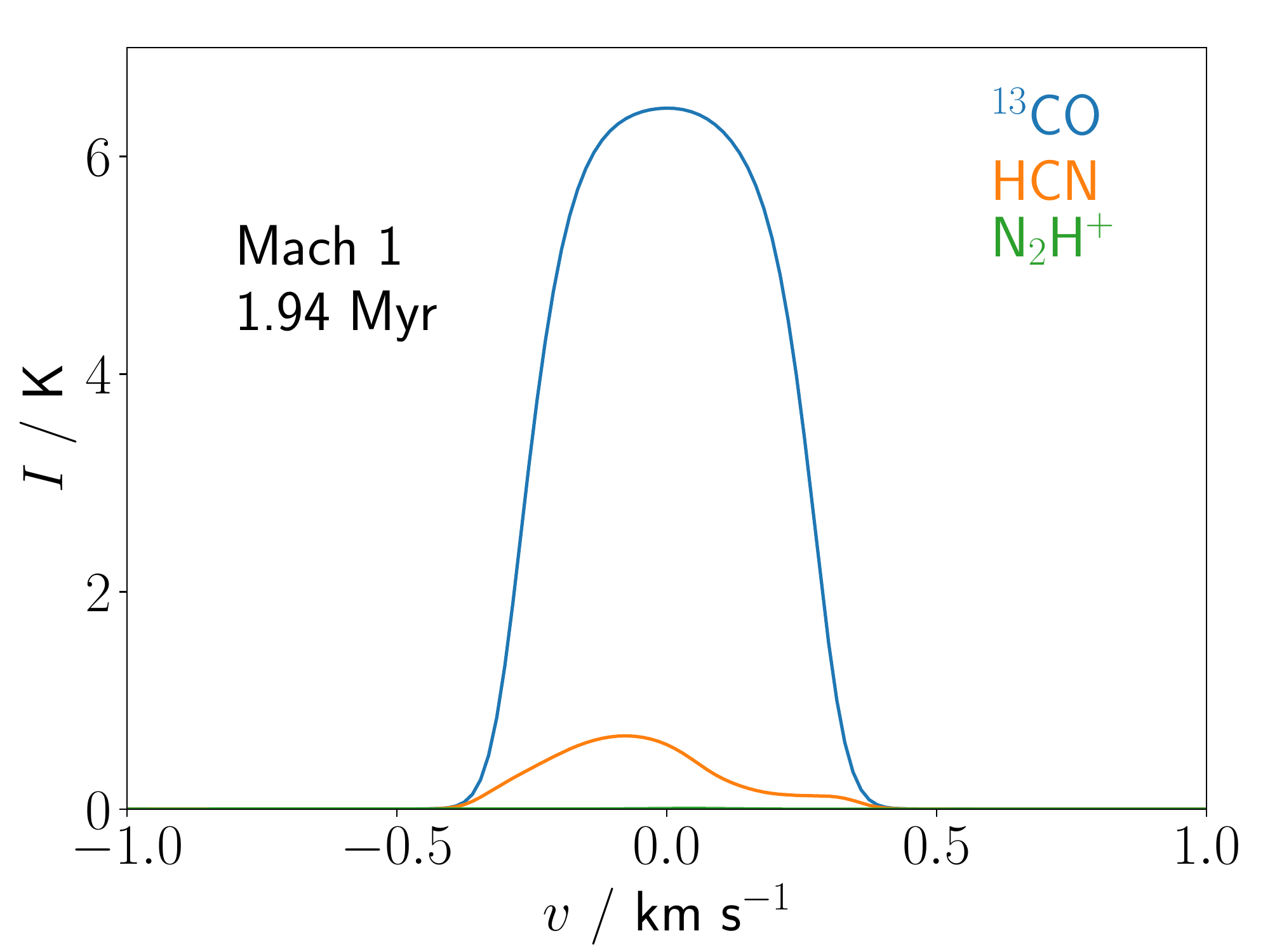}\quad
  \includegraphics[width=0.32\textwidth]{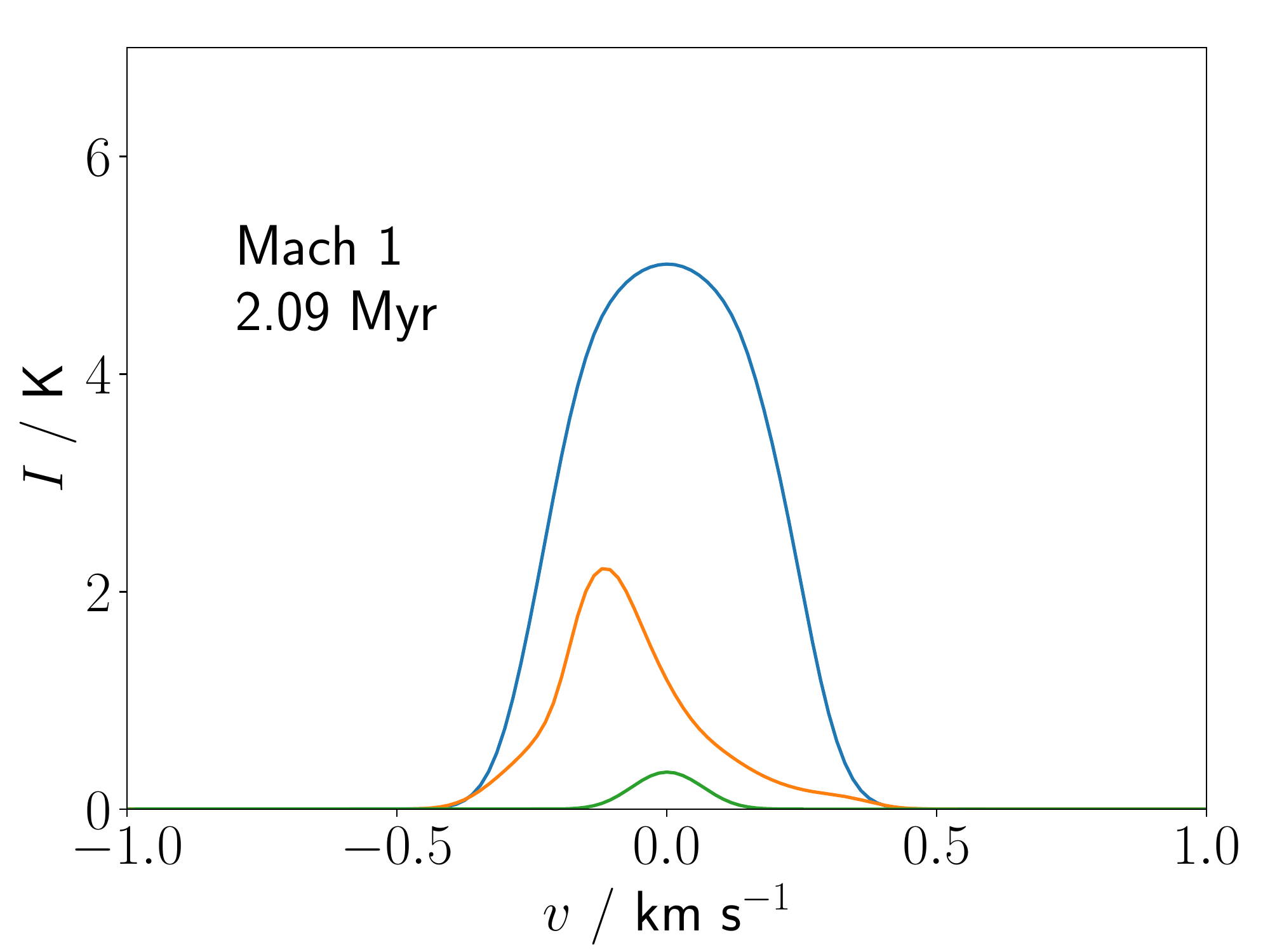}\quad
  \includegraphics[width=0.32\textwidth]{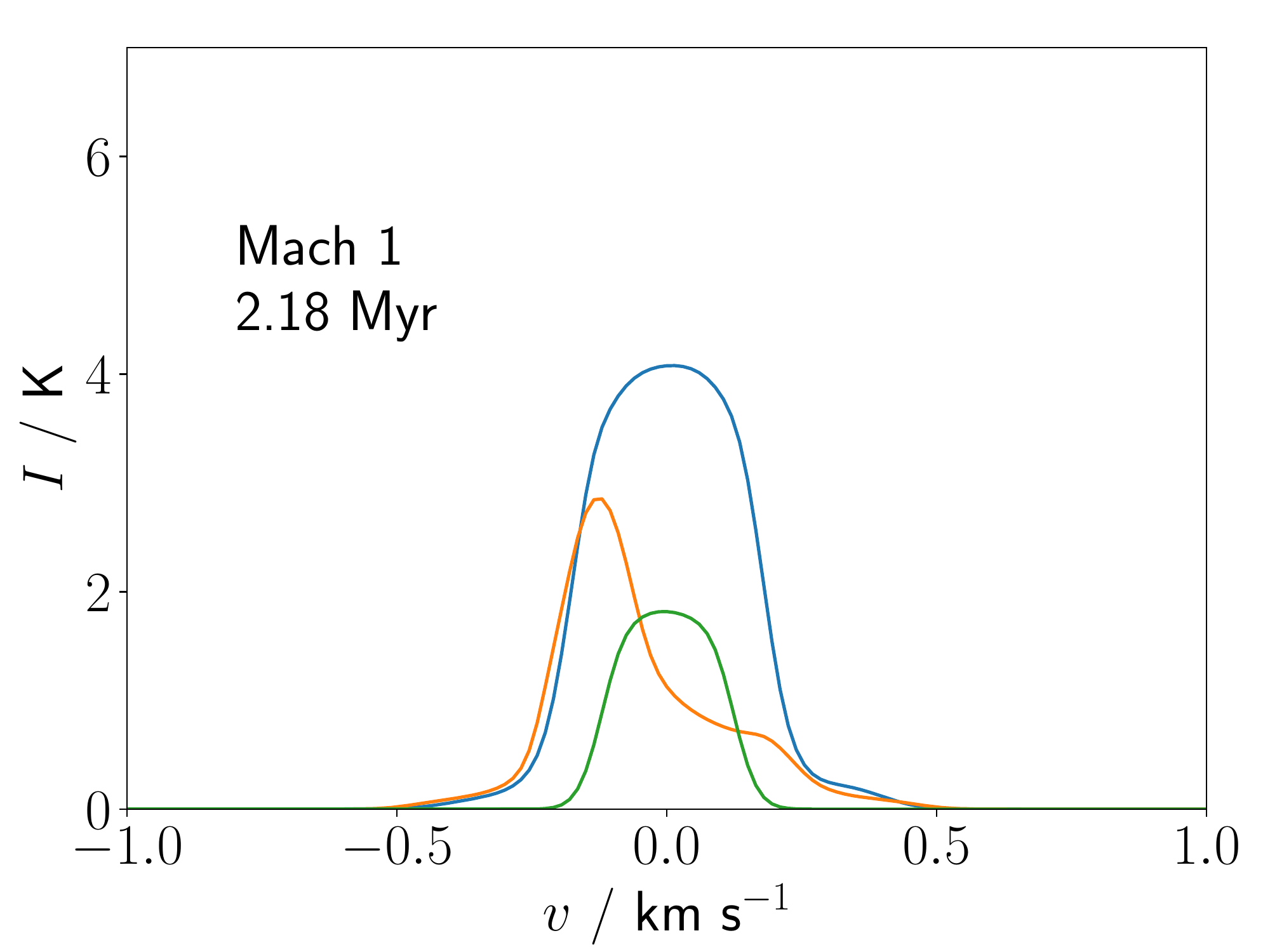}\\
  \includegraphics[width=0.32\textwidth]{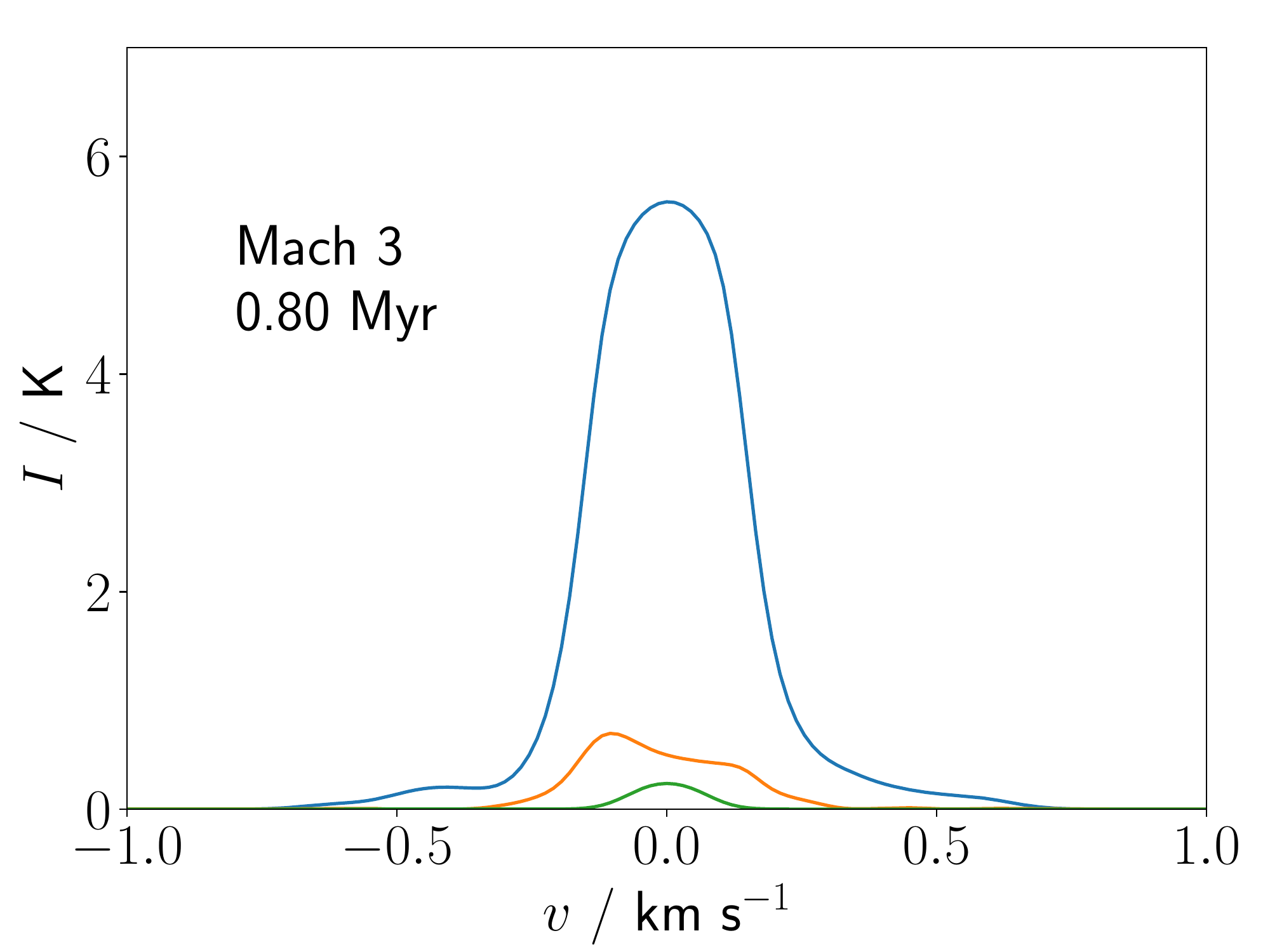}\quad
  \includegraphics[width=0.32\textwidth]{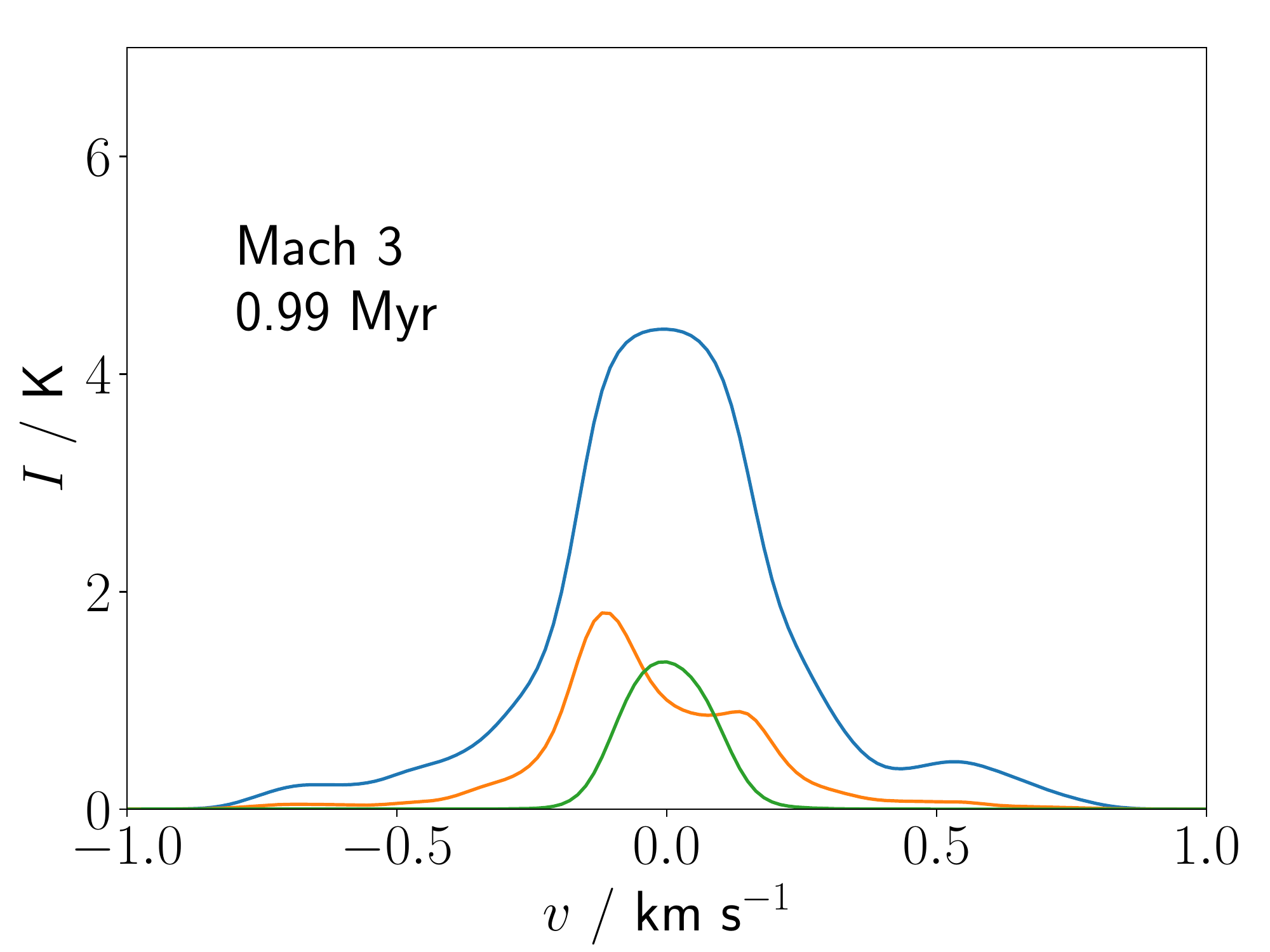}\quad
  \includegraphics[width=0.32\textwidth]{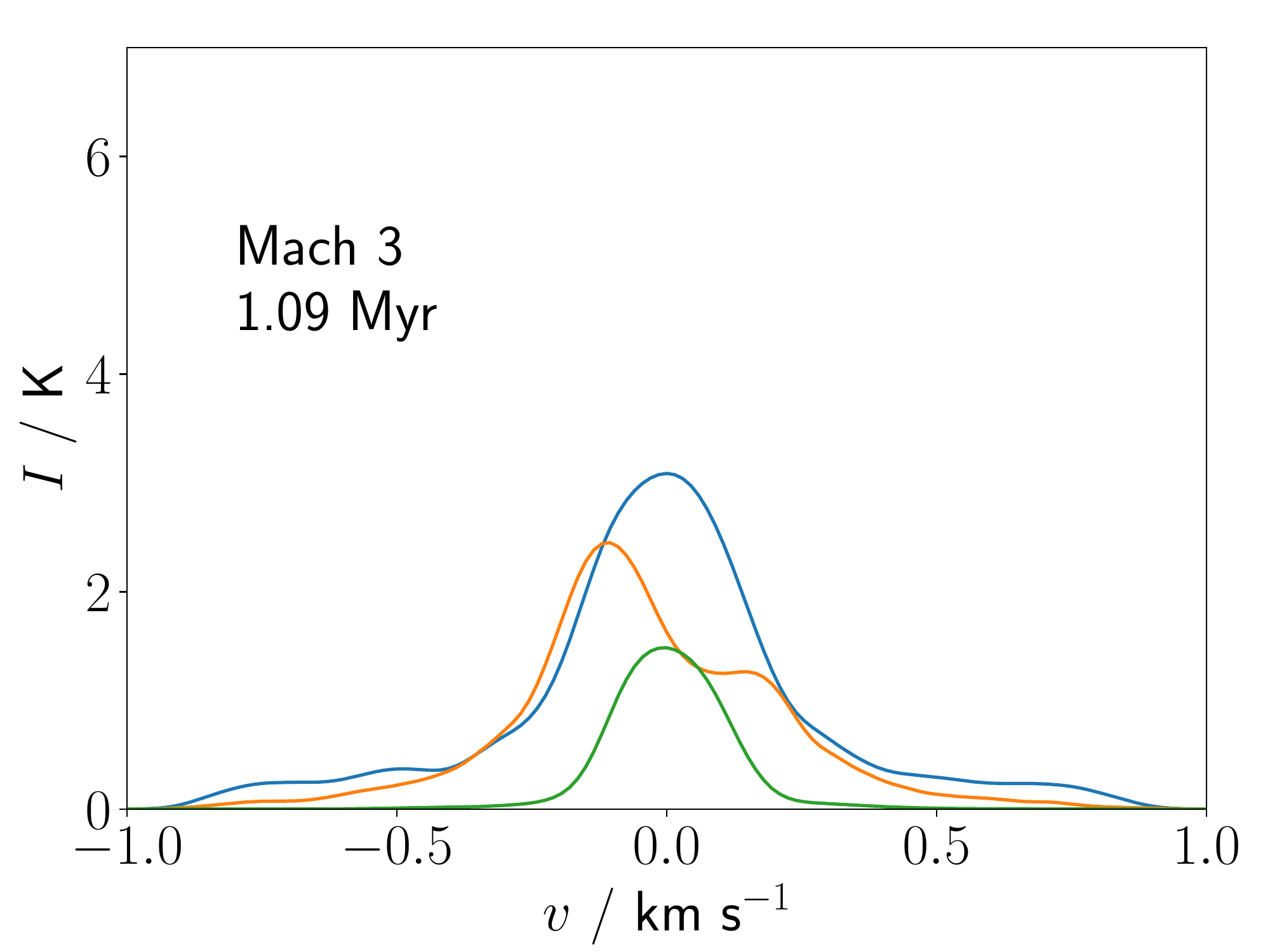}\\
  \caption{Line profiles of the $J=1-0$ rotational transitions of $^{13}$CO (blue), HCN (orange) and N$_2$H$^+$ (green), for ${\cal M} = 1$ (top row) and $3$ (bottom row), evolving with time/central density from left to right.}
  \label{fig:lines}
\end{figure*}

\begin{figure*}
  \centering
  \includegraphics[width=\columnwidth]{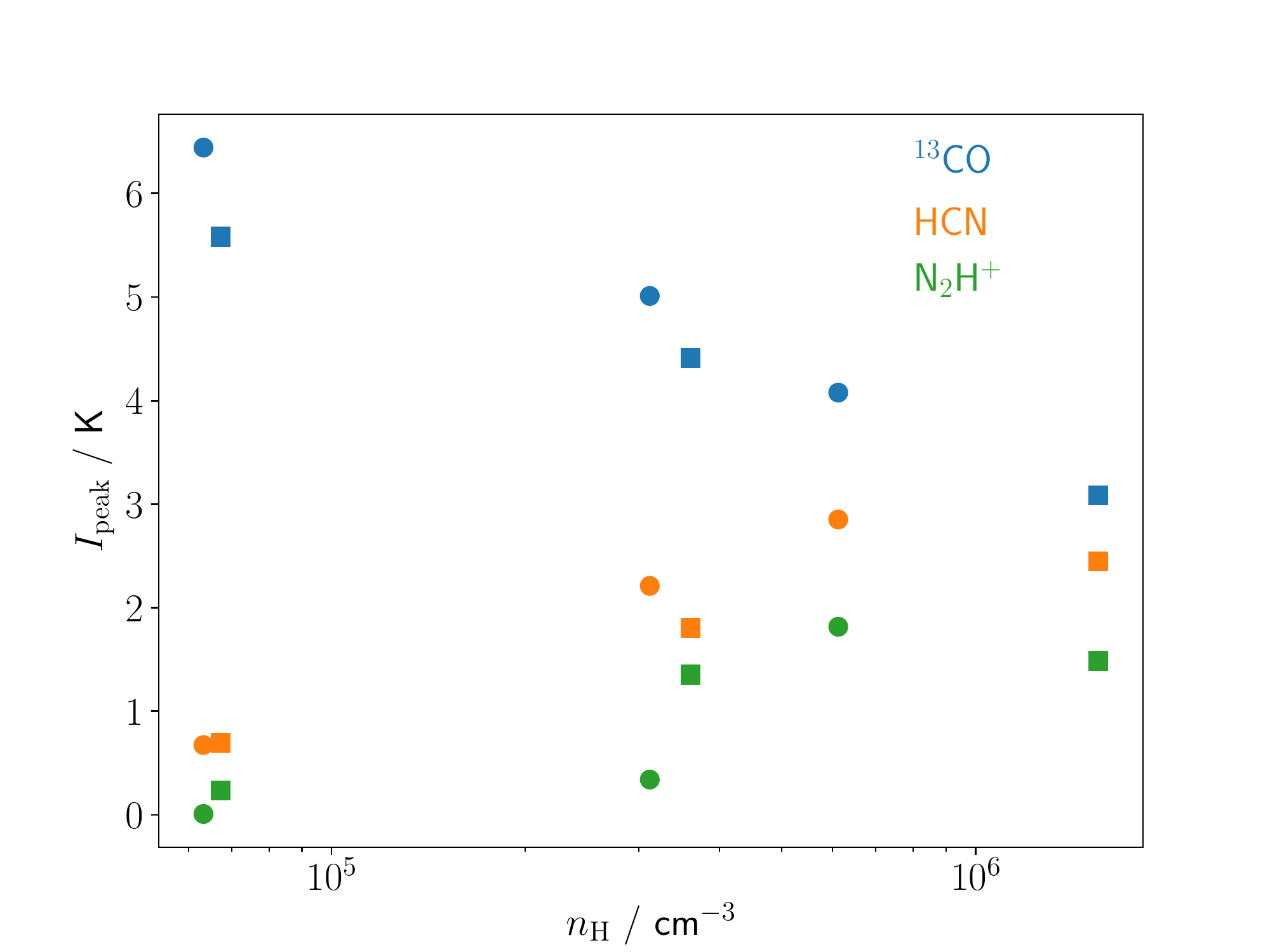}\quad
  \includegraphics[width=\columnwidth]{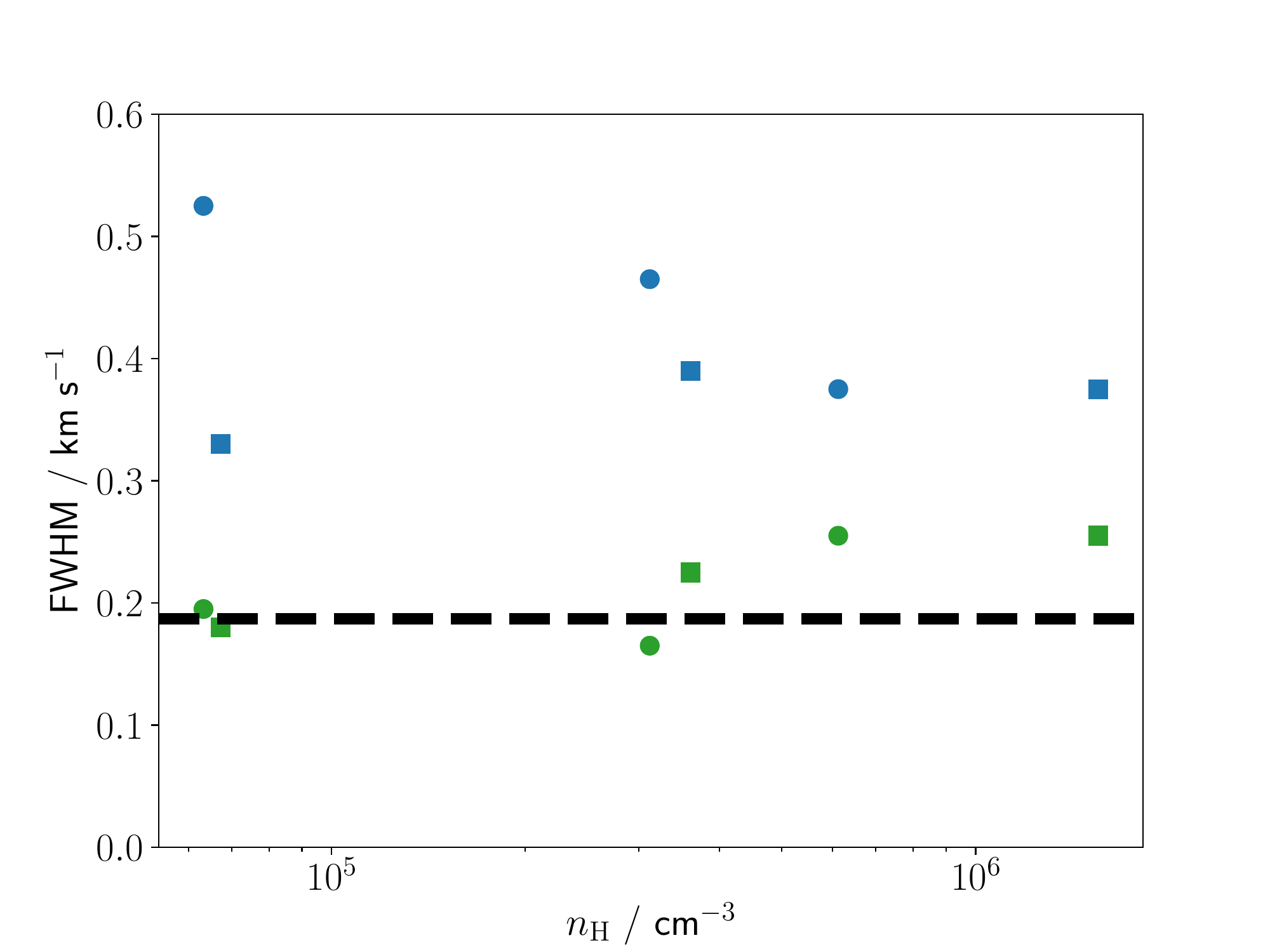}\quad
  \caption{Peak line intensities (left) and FWHMs (right) of the $J=1-0$ rotational transitions of $^{13}$CO (blue), HCN (orange) and N$_2$H$^+$ (green) versus filament central density, for ${\cal M} = 1$ (circles) and $3$ (squares). {HCN widths are not shown, because the asymmetric self-absorbed line profile {(Figure \ref{fig:lines})} makes the FWHM a poor representation of the underlying velocity dispersion.} {The isothermal sound speed for molecular gas at $10 \kel$, $a_0 = 0.187 \kms$, is marked in the right panel by a dashed black line.}}
  \label{fig:densevol}
\end{figure*}

\begin{figure}
  \centering
  \includegraphics[width=\columnwidth]{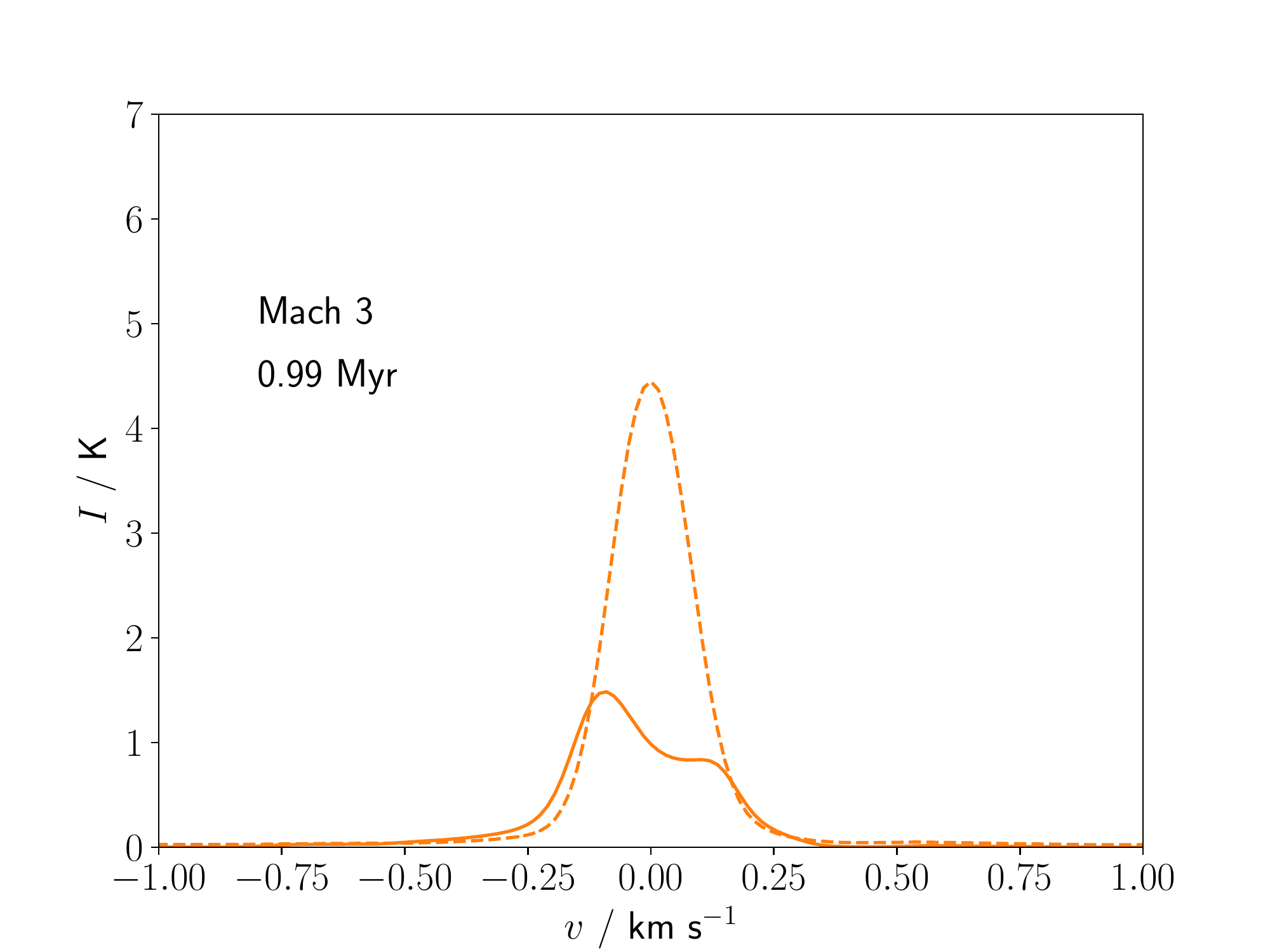}
  \caption{Line profiles of the $J=1-0$ rotational transitions of HCN (solid line) and H$^{13}$CN (dashed line), which has been scaled up by the assumed HCN/H$^{13}$CN abundance ratio of 77 so that for optically thin emission, the line profiles would be identical.}
  \label{fig:optthin}
\end{figure}

\begin{figure*}
  \centering
  \includegraphics[width=\columnwidth]{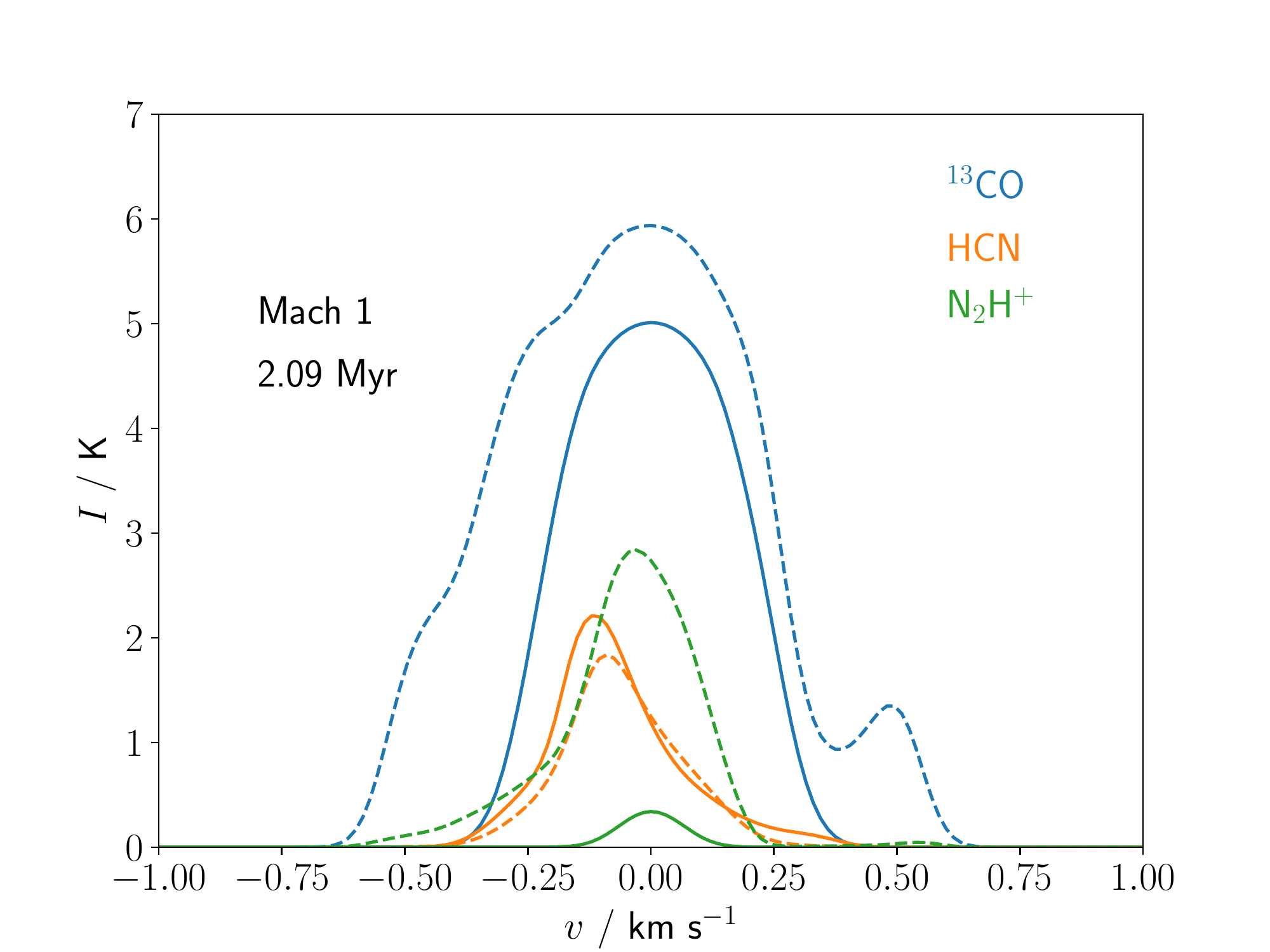}\quad
  \includegraphics[width=\columnwidth]{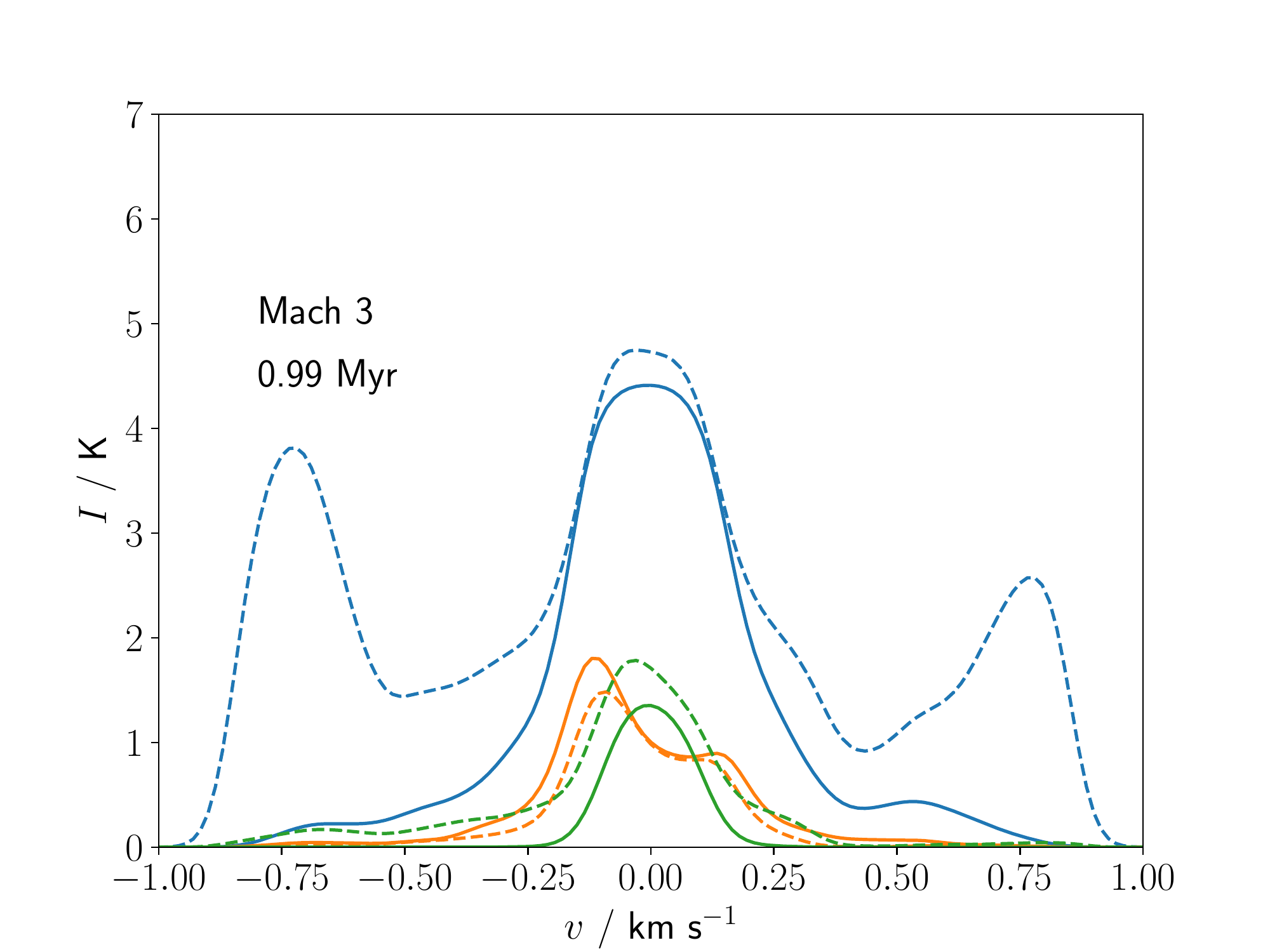}\quad
  \caption{Line profiles of the $J=1-0$ rotational transitions of $^{13}$CO (blue), HCN (orange) and N$_2$H$^+$ (green), for ${\cal M} = 1$ (left) and $3$ (right), and an external UV field of $1.7$ (solid lines) or $0$ (dashed lines) \citet{habing1968} units.}
  \label{fig:radtest}
\end{figure*}

\begin{figure*}
  \centering
  \includegraphics[width=0.32\textwidth]{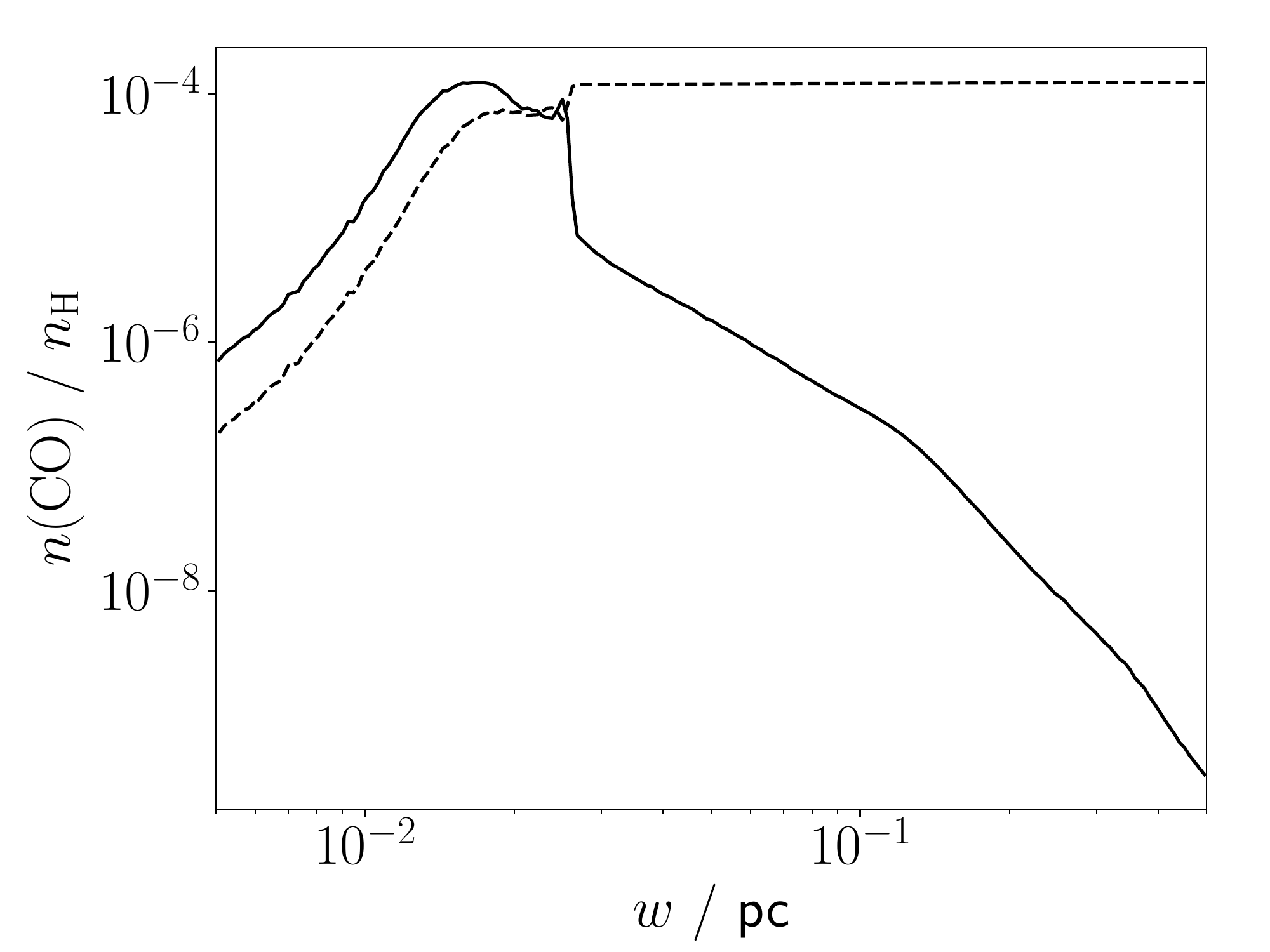}\quad
  \includegraphics[width=0.32\textwidth]{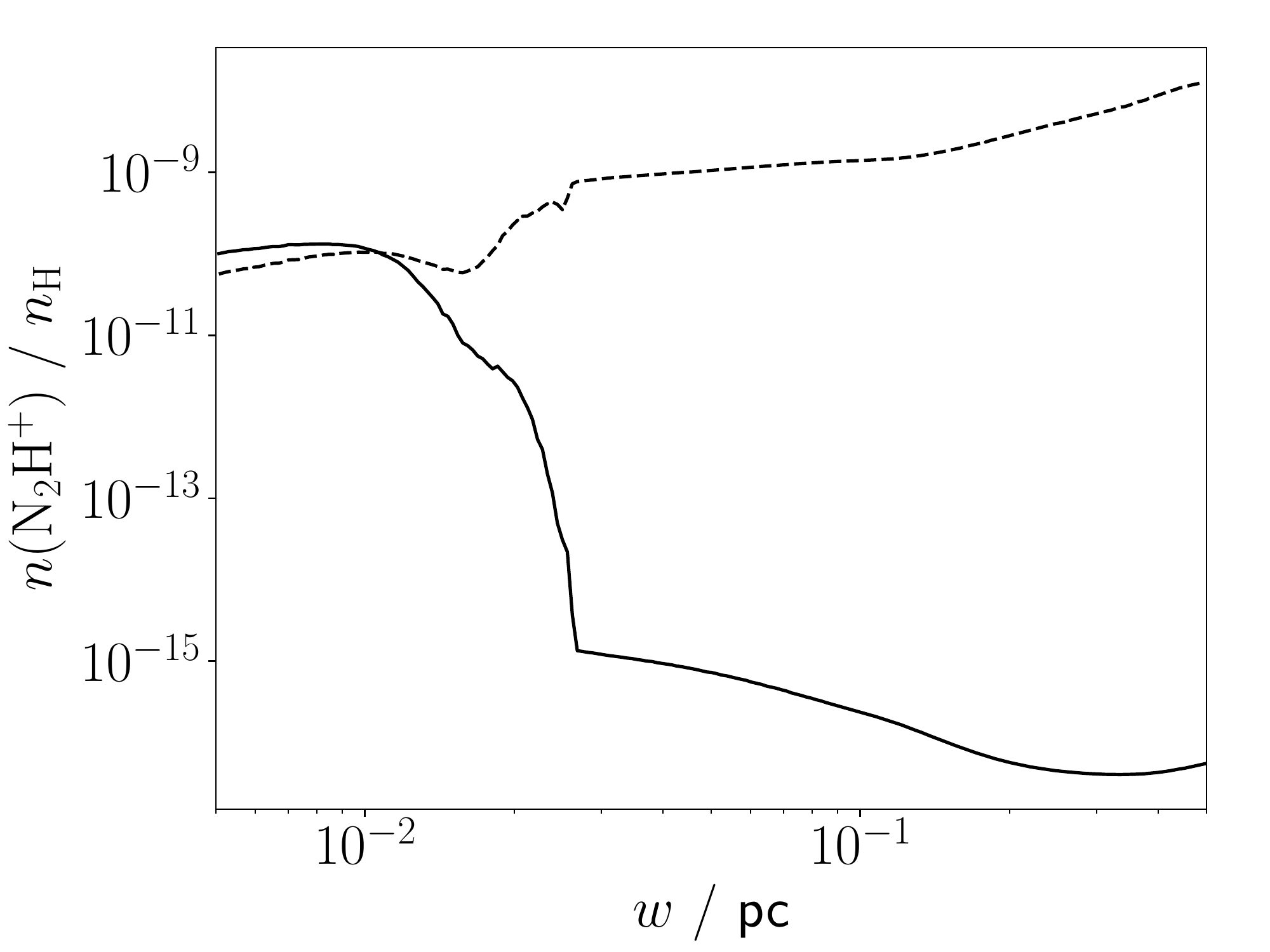}\quad
  \includegraphics[width=0.32\textwidth]{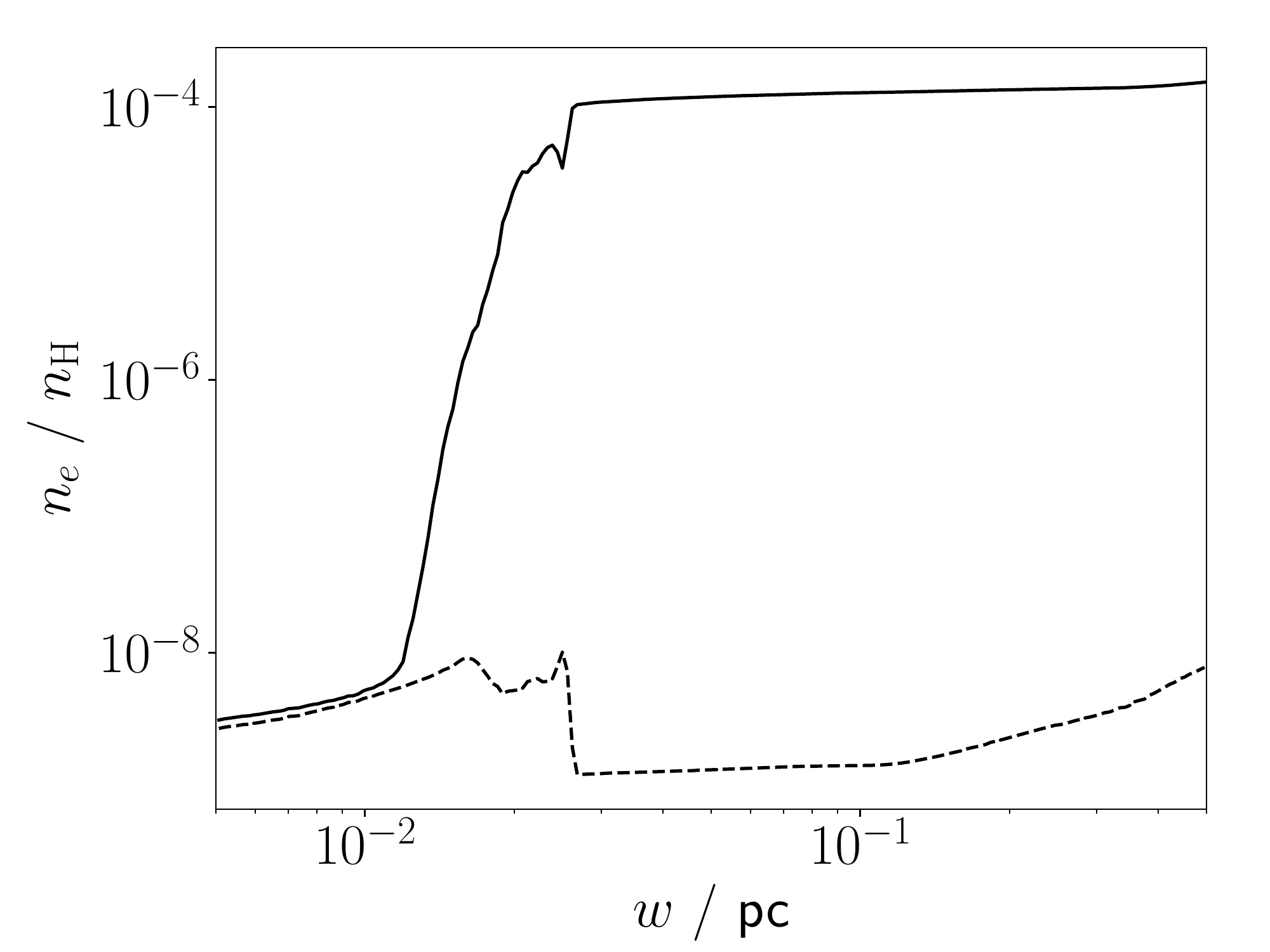}\\
  \caption{{Abundance profiles of CO (left), N$_2$H$^+$ (centre) and e$^-$ (right) for the ${\cal M} = 3$ model at $0.99 \myr$, for an external UV field of $1.7$ (solid lines) or $0$ (dashed lines) \citet{habing1968} units. The radial distance from the axis of symmetry is denoted by $w$.}}
  \label{fig:radabun}
\end{figure*}

\begin{figure*}
  \centering
  \includegraphics[width=\columnwidth]{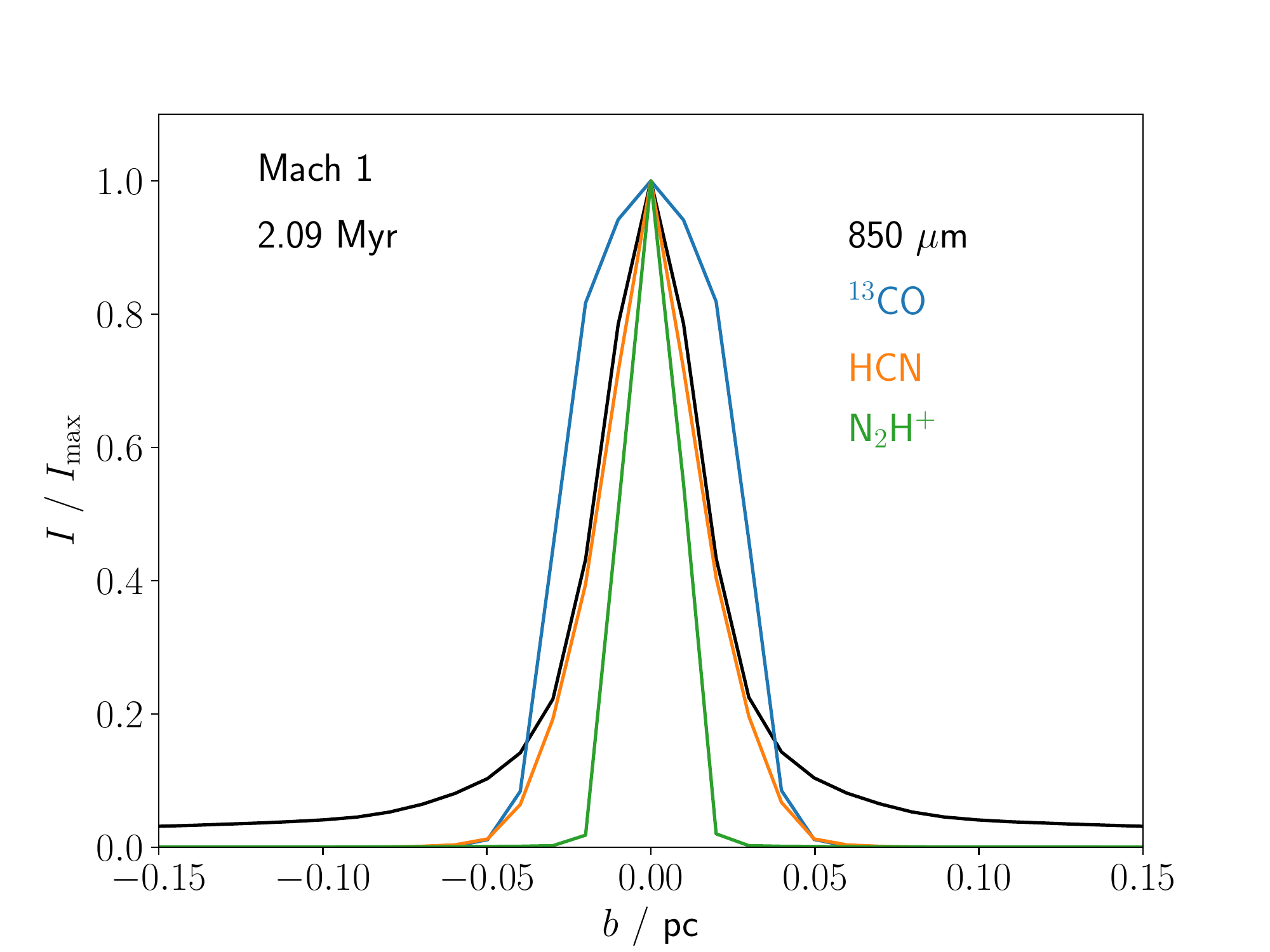}\quad
  \includegraphics[width=\columnwidth]{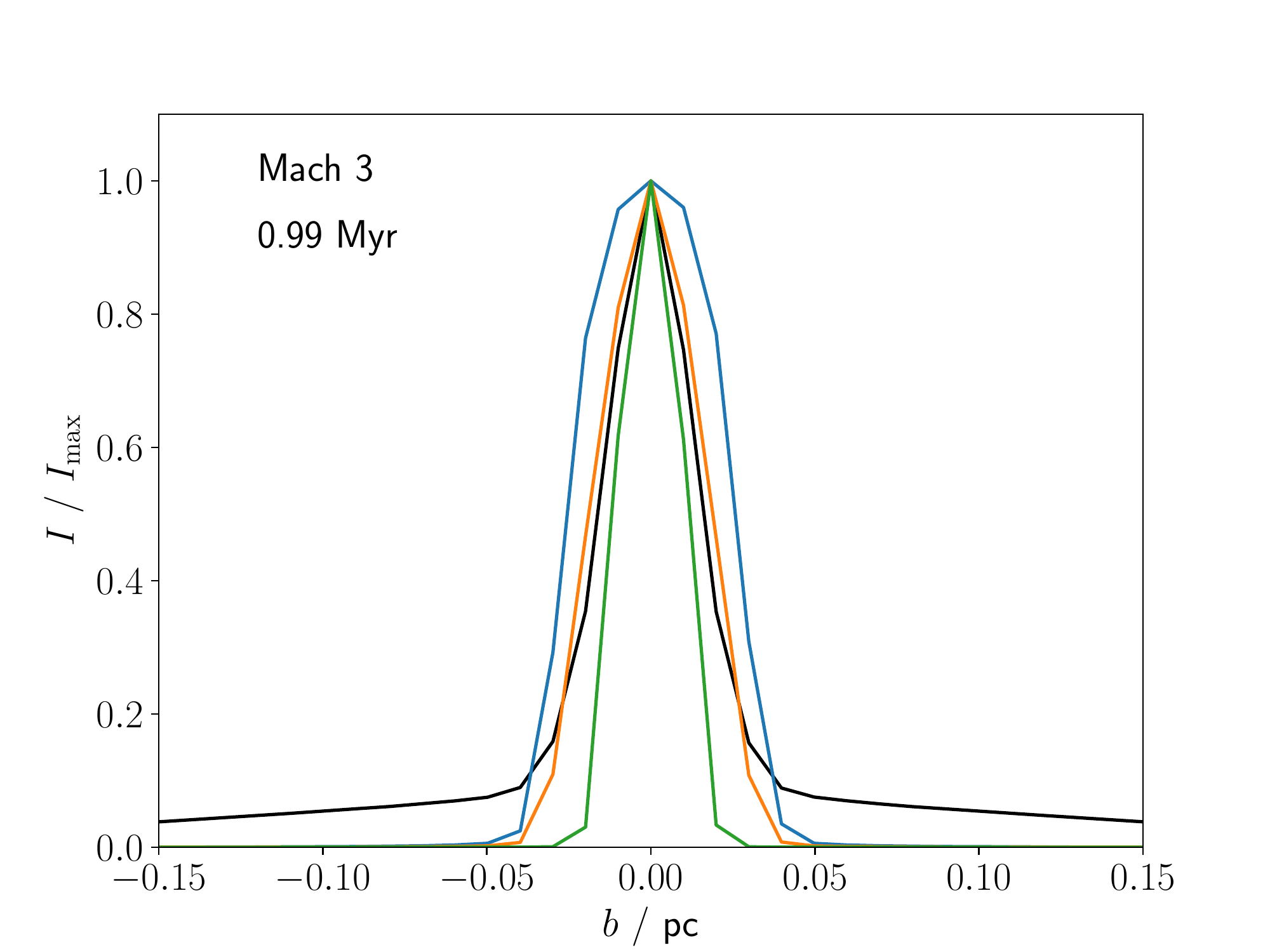}\quad
  \caption{Integrated intensity profiles perpendicular to the filament spine, normalised to the peak value, of the $J=1-0$ rotational transitions of $^{13}$CO (blue), HCN (orange) and N$_2$H$^+$ (green), and of the $850 \um$ thermal dust emission (black), for ${\cal M} = 1$ (left) and $3$ (right). {The projected distance from the axis of symmetry is denoted by $b$.}}
  \label{fig:width}
\end{figure*}

\section{Results}

{Figure \ref{fig:profile} shows the density and radial velocity profiles\footnote{{These profiles are averaged over the $z$ axis between $\pm 1 \pc$, but differences between the average profiles and individual cuts are minor \citep{priestley2022}. }} for the ${\cal M} = 1$ and ${\cal M} = 3$ models at three epochs, chosen so that the corresponding model with a different ${\cal M}$ value has a comparable central density. As discussed in \citet{priestley2022}, the gradient of the density profiles undergoes an abrupt steepening at the location of the accretion shock, clearly visible in the velocity profiles as a sharp transition from supersonic to subsonic inflow. This serves as a useful distinction between the central filament and the ambient material {out of which the filament has formed}. The location of the filament boundary moves inwards with time, while the peak inflow velocity of the ambient material increases due to the gravitational attraction of the central filament. Filaments with ${\cal M} = 1$ are broader than their ${\cal M} = 3$ counterparts at equal central density due to the lower ram pressure of the accretion flow; {the density profiles are not solely set by self-gravity, resulting in shallower slopes than the hydrostatic equilibrium solution \citep{priestley2022}, as is observed \citep{palmeirim2013}}.}

Figure \ref{fig:lines} shows the $^{13}$CO, HCN and N$_2$H$^+$ line profiles\footnote{{We average the line profiles within a radius of $0.05 \pc$ of the origin, approximating the effect of typical single-dish beam sizes for nearby ($\sim 100 \pc$) filaments. As the line emission is dominated by the central filament, increasing the averaging radius has a negligible effect on the shape of the profiles, although the averaged intensities are lower.}} at different evolutionary stages. The $^{13}$CO line is stronger and broader than the others, and is typically symmetric, although in some cases `wings' around the central line corresponding to the inflowing material are visible. HCN profiles are asymmetric, showing the characteristic infall signature of redshifted self-absorption. However, this self-absorption is not due to the inflowing material, which can also be seen as extended wings in some cases. The velocity of the self-absorption is much lower than {that of the} inflow, and arises instead from the gravitational contraction of the dense central filament. N$_2$H$^+$ profiles are symmetric, showing no self-absorption, and have negligible emission from the inflowing material.

Figure \ref{fig:densevol} shows the evolution of line peak intensities and full-widths at half maximum (FWHMs) with central density. Line properties are similar {between the two values of ${\cal M}$} when the central densities are comparable, making this a much better proxy for evolutionary stage than the age; the ${\cal M} = 3$ model terminates (defined as the point at which it begins to fragment along its longitudinal axis; \citealt{priestley2022}) before the ${\cal M} = 1$ model shows any detectable emission from lines other than $^{13}$CO. The peak intensity of the $^{13}$CO line decreases as the filament evolves, due to depletion of the molecule onto grain surfaces as the density rises. Peak intensities of both the HCN and N$_2$H$^+$ lines increase {monotonically with central density}. This suggests that the ratio of the peak intensity of lines tracing denser gas to those of CO isotopologues may serve as an evolutionary indicator, with higher values indicating relatively more evolved filaments, although the normalisation of this indicator is likely to be too sensitive to the local conditions to accurately diagnose the central filament volume density. {The peak intensities of the HCN and N$_2$H$^+$ lines will also depend on which hyperfine component is being measured (Appendix \ref{sec:hfs}), but in all cases should still increase relative to $^{13}$CO with evolutionary stage.}

Line FWHMs, by contrast, show relatively little evolution, being sub- to transonic throughout the filament lifetime, as observed {for filaments with ${\cal G} \sim 1$} \citep{arzoumanian2013,hacar2013}.\footnote{{\citet{arzoumanian2013} find that filaments with line densities significantly in excess of $\mu_{\rm CRIT}$ tend to have higher velocity dispersions, indicating supersonic non-thermal motions. This may be a sign of rapid infall, caused by the much greater importance of self-gravity in these objects.}} Notably, $^{13}$CO FWHMs from the ${\cal M} = 1$ model are generally higher than those of the ${\cal M} = 3$ model, despite the lower inflow velocity. This is most apparent at the lowest central densities ($\sim 6 \times 10^4 \pcc$), where the ${\cal M} = 3$ $^{13}$CO FWHM is $\sim 0.2 \kms$ below the equivalent ${\cal M} = 1$ value. This appears to be due to blending of the emission from the central filament and the inflowing material, {as the ${\cal M} = 1$ inflow velocities only shift the line centroids by approximately one thermal linewidth}. {This effect, combined with additional opacity broadening \citep{hacar2016b}, makes the $^{13}$CO linewidth a poor tracer of the velocity dispersion within the filament, which it will inevitably overestimate. The N$_2$H$^+$ line, with a much lower opacity and a negligible contribution from the inflowing material, is more representative of the conditions in the central filament, which from Figure \ref{fig:profile} are clearly subsonic at all times.} {\citet{smith2012} obtained similar results for the N$_2$H$^+$ line from studying filaments formed in a larger-scale simulation of a turbulent molecular cloud, suggesting that subsonic line widths occur generically, even in scenarios less idealised than our models.}

{We do not show HCN FWHMs in Figure \ref{fig:densevol} because the self-absorbed line profile makes this an unrealiable tracer of the intrinsic line width. Figure \ref{fig:optthin} compares the HCN line profile to that of H$^{13}$CN, an optically-thin isotopologue.\footnote{{Collisional data for H$^{13}$CN are not available, so we assume the same properties as HCN, with the only difference being the molecular abundance (reduced by a factor of 77 for H$^{13}$CN).}} As with N$_2$H$^+$, the intrinsic velocity dispersion of the line, as represented by that of the symmetric H$^{13}$CN profile, is clearly subsonic.}

As noted above, we have assumed what is effectively a lower limit for the column density shielding the filament material from any background UV field. Figure \ref{fig:radtest} compares these results with line profiles resulting from the extreme case of no background UV field whatsoever. This results in stronger N$_2$H$^+$ emission, particularly for the ${\cal M} = 1$ model, {but line widths are not greatly affected, although this is difficult to quantify due to the asymmetry of the no-UV profiles}. {This increase in intensity is due to the main destruction mechanism outside the filament centre changing from dissociative recombination with free electrons to reactions with CO. Figure \ref{fig:radabun} shows the abundance profiles of the relevant species for the ${\cal M} = 3$ model. With no external UV field, almost all carbon is in the form of CO at all radii, whereas a non-zero UV flux produces an extensive region of photoionised C$^+$ with a much higher electron density. The dissociative recombination rate for N$_2$H$^+$ is significantly higher than that of reactions with CO \citep{mcelroy2013}, so even a moderate electron density causes a sharp drop in the N$_2$H$^+$ abundance. With no UV field, the abundance of the main source of destruction is almost unchanged ($\sim 10^{-4}$), but the reaction rate is much lower, and N$_2$H$^+$ maintains a roughly constant abundance throughout the filament, although the densities at large radii are too low to excite detectable line emission.}

The $^{13}$CO line profiles, on the other hand, are significantly altered {by the removal of an external UV field}; without photodissociation of the molecule in the lower-density accreting material, {the greatly increased abundance (Figure \ref{fig:radabun}) causes} additional peaks to appear at velocities corresponding to the initial inflow, which are blended with the central filament emission for the ${\cal M} = 1$ model. These predicted line properties (multiple peaks and/or substantially broadened emission) are quite unlike those observed for CO isotopologues \citep{arzoumanian2013,panopoulou2014,hacar2016,suri2019}. {If our model is applicable to real filaments, this suggests that they cannot be so deeply embedded in ambient material as to be substantially shielded from the background UV field.}

{The sample of filaments in \citet{arzoumanian2019} has a range of background column densities spanning $1-10 \times 10^{21} \pcs$, which corresponds to extinctions associated with the parent cloud material of $0.6-6 \, {\rm mag}$ \citep{bohlin1978}. However, the relevant quantity for shielding is the {\it effective} $\av$, a weighted average over all lines of sight from the filament itself \citep{glover2007}. This quantity is typically lower than the line-of-sight $\av$, because low-$\av$ sightlines contribute disproportionately to the total photodissociation rate. \citet{clark2014} find that observed extinctions of up to $20-30$ mag only correspond to $\sim 2-3$ mag of effective shielding, so even the highest background values from \citet{arzoumanian2019} are still consistent with the filaments being at most moderately shielded ($\av \lesssim 1 \, {\rm mag}$) from the external UV field, as we require. In regions of active star formation, where the UV strength may be significantly higher than the Solar neighbourhood value adopted here, an even higher effective $\av$ would be required to reach a situation comparable to the no-UV case in Figure \ref{fig:radtest}.}

\section{Discussion}

\subsection{Signatures of filament accretion}

Our results suggest that in a scenario where filaments form via converging turbulent flows, these accretion flows would be difficult to detect observationally, even for supersonic velocities. The accreting material is only apparent as relatively faint wings around the central line profile, even for low-density tracers such as $^{13}$CO, unless this material is well-shielded from the ambient UV field. Line widths are at most transonic, and do not correspond to the original {inflow} velocity. Where lines show self-absorption features commonly associated with inflows, such as the HCN profiles in Figure \ref{fig:lines}, these are associated with the central filament itself and not the material accreting onto it. Distinct velocity components in lines other than those from CO isotopologues, such as those in HC$_5$N identified by \citet{smith2023}, are likely to represent distinct dense filamentary structures, rather than accretion onto a main filament. {However, we note that this only holds when the accreting material has a relatively low density, rather than itself being shock-compressed (see Section \ref{sec:sheet}).}

\subsection{Fragmentation and substructure}

Our simulations are not designed to study substructure; they have cylindrical symmetry, and are terminated before any longitudinal fragmentation can occur. Nonetheless, the fact that the N$_2$H$^+$ $J=1-0$ line almost exclusively traces the dense filament material may have some implications for the `fibres' that are identified in PPV space within observed filaments \citep{hacar2013,hacar2018,shimajiri2019b}. Previous theoretical studies of filaments with substructure have focused on transitions from CO isotopologues \citep{seifried2017a,clarke2018}, which likely includes emission from material which has not yet been incorporated into a filament (or fibre). This may account for the lack of correspondance between PPV structures, identified via CO lines, and actual physical density structures \citep{clarke2018}; lower-density material has a disproportionate contribution to the observed line intensity relative to its mass. N$_2$H$^+$ has the opposite behaviour \citep{tafalla2021}, preferentially highlighting the densest material. Substructures identified via this line, and others with similar behaviour, are more likely to represent genuinely separate physical entities.

\subsection{Line widths in filaments and cores}

Despite forming from supersonic turbulent motions, even our ${\cal M} = 3$ model displays at most transonic line widths, in agreement with observations {of filaments with comparable (${\cal G} \sim 1$) line densities}. Similarly to filaments, prestellar cores are generally observed to have sub- to transonic line widths \citep[e.g.][]{lee1999,tafalla2002,li2023}, {but models of isolated cores almost inevitably develop supersonic infall motions \citep{larson1969} which are predicted to be readily apparent in the molecular line profiles \citep{yin2021,priestley2022b}.\footnote{{Models with high magnetic field strengths predict line profiles in good agreement with observations, but magnetic fields in prestellar cores are typically found to be much weaker than the required values \citep{crutcher2012,pattle2022}}} This conflict is often resolved via} highly extended initial density profiles \citep{keto2015,sipila2018,sipila2022}, {which do not reach supersonic infall velocities before the formation of a protostar, but these structures} are well beyond the {limit for gravitational stability,} and so represent quite implausible initial conditions to find in a turbulent molecular cloud \citep{whitworth1996}. Our results, {like those of \citet{smith2012}}, suggest that structures formed from supersonic flows tend to efficiently dissipate the original velocity, and that the subsequent gravitational contraction has a much lower velocity than would be expected if the object were initially at rest. This {may} provide an alternative explanation for the {narrow} observed line widths in prestellar cores.

\subsection{Limitations of cylindrical symmetry}
\label{sec:sheet}

{Filaments are often found to have transverse velocity gradients \citep{arzoumanian2018,shimajiri2019,watkins2019,chen2022,kim2022}. While this is sometimes attributed to rotation around the long axis \citep[e.g.][]{alvarez2021}, a more common interpretation is that the gradient represents material being accreted from a sheet-like structure at an inclination to the line of sight \citep{pineda2022}, so that lines are redshifted on one side of the filament and blueshifted on the other. The sheet itself is often suggested to have formed in the interface between two colliding molecular clouds \citep[e.g.][]{balfour2015,balfour2017,inoue2018,abe2021}. As our models are cylindrically symmetric by construction, they are incapable of reproducing these gradients, which would require angular variation in the properties of the accreting gas. However, synthetic line observations of turbulent molecular clouds \citep{priestley2020} find that the resulting filamentary structures do in fact tend to have velocity gradients perpendicular to their spines, even in the case of isolated clouds with no collision-induced layer. This suggests that supersonic turbulence by itself is capable of producing the sheet-like structures inferred observationally, particularly as some degree of compression is necessary in order for the accreting material to be detectable in line emission.}

\subsection{Filament widths in different tracers}

{It is expected theoretically \citep{priestley2020} and found observationally \citep{shimajiri2023} that the same filament will appear to have different widths when observed in different tracers. This is due to the complex combination of excitation, abundance and optical depth effects which relate the column density along the line of sight to the observed line intensity. Figure \ref{fig:width} shows the spatial variation of the integrated line intensities perpendicular to the filament spine. As the spatial resolution of our radiative transfer models ($0.01 \pc$) is much worse than that of the underlying SPH simulation, we compare the line data to $850 \um$ dust continuum emission images produced with {\sc lime}. These have identical spatial resolution to the line PPV cubes, and should trace the column density almost exactly, as we assume a constant dust temperature of $10 \kel$ and the optical depth at this wavelength is extremely low ($\tau_{850} < 10^{-4}$). The filaments appear to be broader than the underlying density profile when observed in $^{13}$CO, and narrower when observed in N$_2$H$^+$. This likely explains the different width distributions found by studies in these two tracers \citep{panopoulou2014,hacar2018} when compared to those using dust continuum emission \citep{arzoumanian2011,arzoumanian2019}.}

{The different widths can be understood as a consequence of the relationship between intensity and column density of the two lines \citep{tafalla2021}; the intensity of CO isotopologues saturates at a relatively low {hydrogen} column density of $N_{\rm H_2} \sim 2 \times 10^{21} \pcs$, whereas the N$_2$H$^+$ line only becomes visible above $10^{23} \pcs$. The $^{13}$CO intensity therefore overrepresents the lower-density outer parts of the filament, whereas the N$_2$H$^+$ line mainly traces the densest central regions. Interestingly, the HCN line, which \citet{tafalla2021} find to have a linear relationship between intensity and column density, does in fact appear to accurately trace the shape of the underlying density profile, at least in the central part of the filament.}

\subsection{Preexisting versus dynamical filaments}

It is common practice in theoretical studies to assume that a filament already exists at the outset of the model, in that the initial radial density profile is quite centrally condensed \citep[e.g.][]{seifried2015,seifried2016,gutierrez2023,misugi2023}. If filaments are actually structures formed via turbulent accretion, this is inappropriate. Even if the initial filament also has supersonic converging motions \citep[e.g.][]{clarke2016}, there is no guarantee that the resulting density and velocity profiles will resemble those in the more realistic scenario where no filament exists at the outset, particularly given that the definition of what counts as a filament seems to be strongly related to the accretion shock \citep{priestley2022}. We suggest that future studies should focus on forming the objects of interest self-consistently \citep[e.g.][]{smith2012,smith2013}, in keeping with their presumed origins from cloud-scale turbulent motions \citep{maclow2004}.

\section{Conclusions}

The filaments identified in molecular clouds are thought to form due to supersonic turbulent motions within these larger objects. We investigate the observational consequences of this scenario, specifically the predicted molecular line emission properties of filaments formed via supersonic flows. These are largely consistent with data; in particular, line widths are trans- to subsonic, even for highly supersonic inflows. The central filament dominates the line emission, making it difficult to identify signs of the accreting material. {Signatures of infall, such as redshifted self-absorption in the HCN $J=1-0$ line, are due to the contraction of the filament itself, rather than the accretion flows it formed out of.} Additional velocity components would only be visible in the limit of almost total shielding from the external UV field, and even then only for CO isotopologues. Our results support the idea that filaments in molecular clouds are dynamically-forming objects, which cannot be modelled as equilibrium structures, or considered in isolation from their environment.

\section*{Acknowledgements}
We are grateful to Michael Anderson and Nicolas Peretto for their advice on measuring velocity dispersions, {and to the anonymous referee for a useful report}. FDP and APW acknowledge the support of a consolidated grant (ST/K00926/1) from the UK Science and Technology Facilities Council (STFC).

\section*{Data Availability}
The data underlying this article will be shared on request.

\bibliographystyle{mnras}
\bibliography{filchem}

\appendix

\section{Hyperfine structure}
\label{sec:hfs}

\begin{figure*}
  \centering
  \includegraphics[width=\columnwidth]{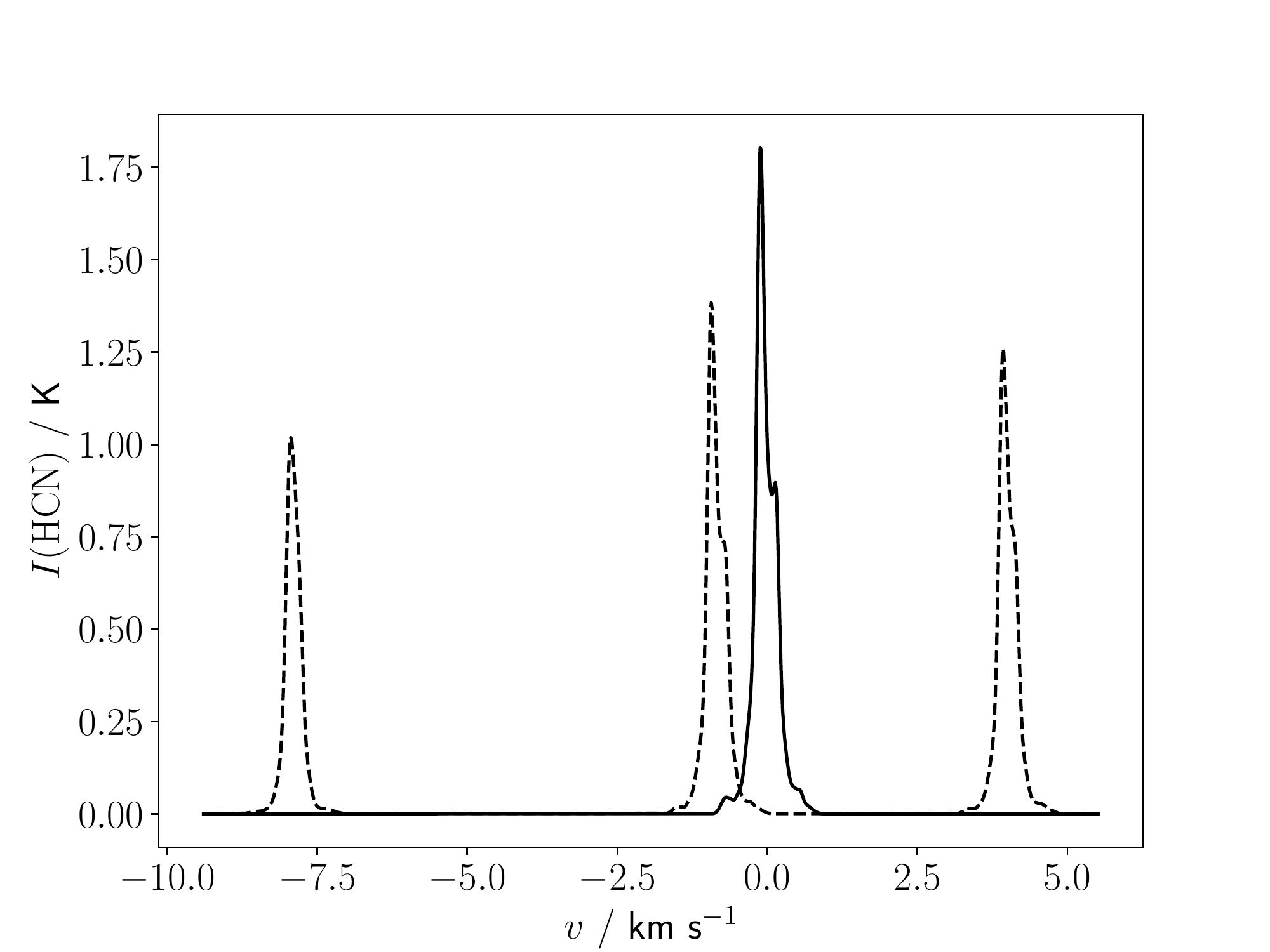}\quad
  \includegraphics[width=\columnwidth]{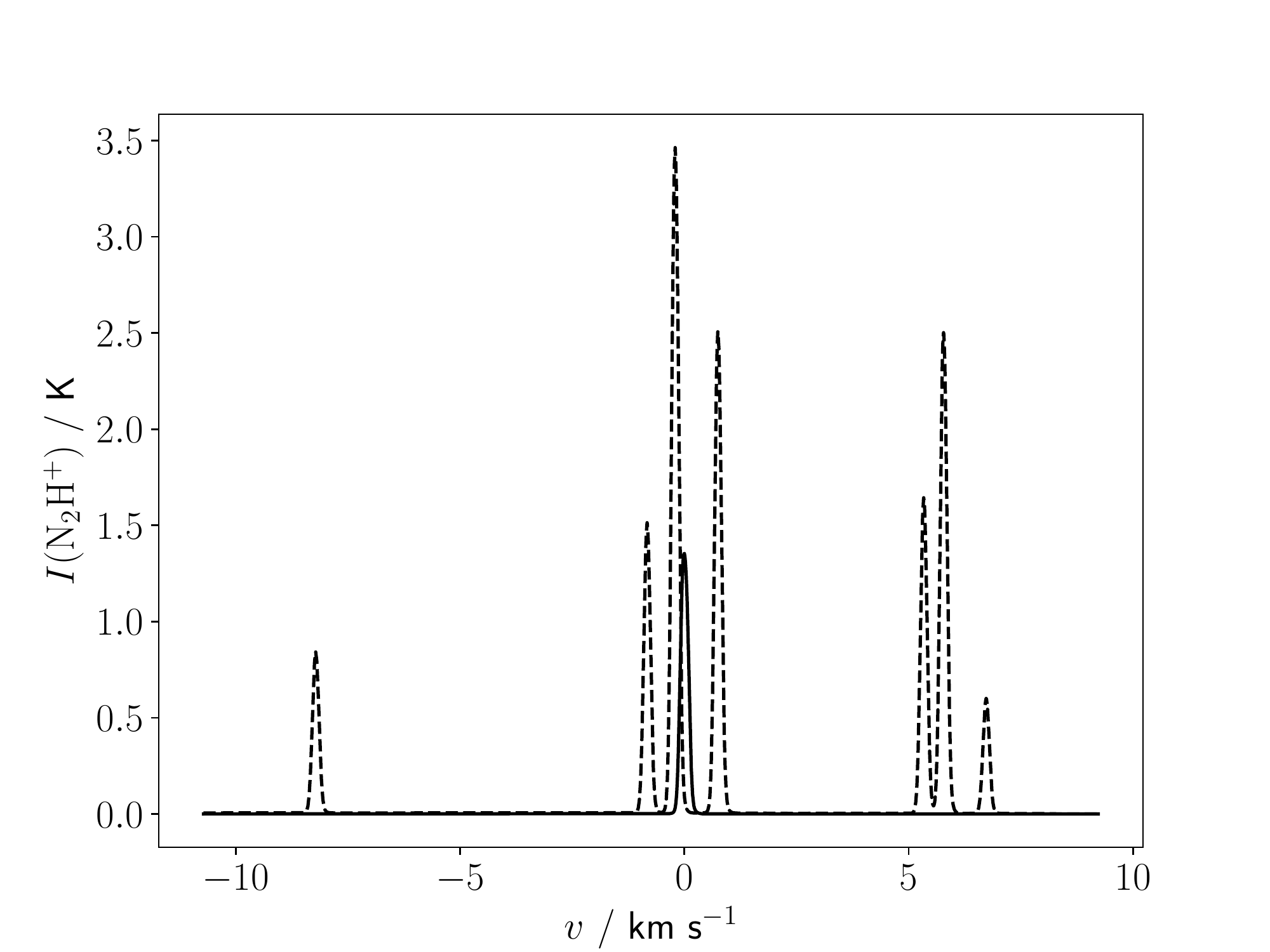}\quad
  \caption{HCN (left) {and N$_2$H$^+$ (right)} $J=1-0$ line profiles for the ${\cal M} = 3$ model at $0.99 \myr$, with (dashed lines) and without (solid lines) hyperfine structure.}
  \label{fig:hfs}
\end{figure*}

\begin{figure*}
  \centering
  \includegraphics[width=\columnwidth]{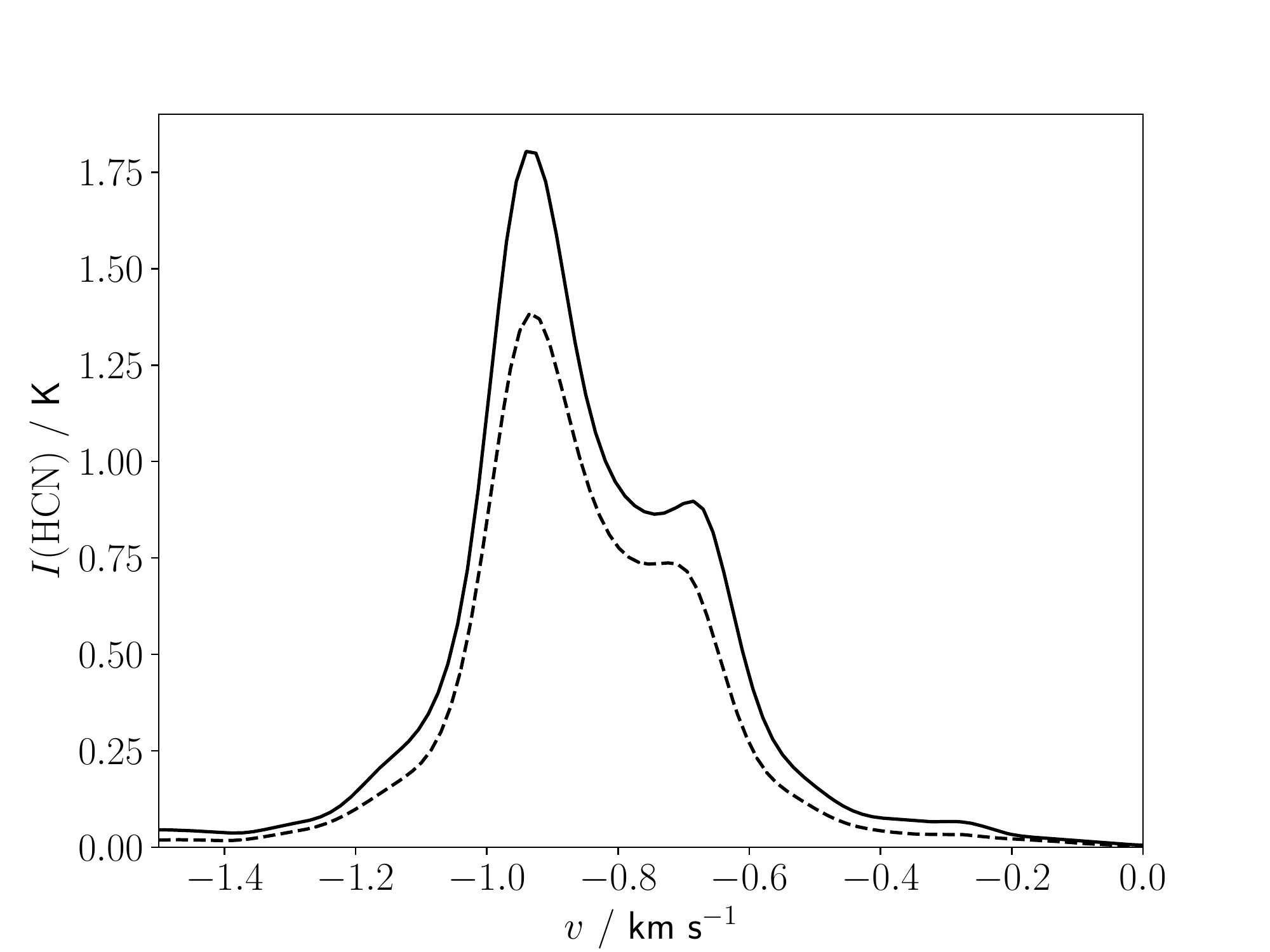}\quad
  \includegraphics[width=\columnwidth]{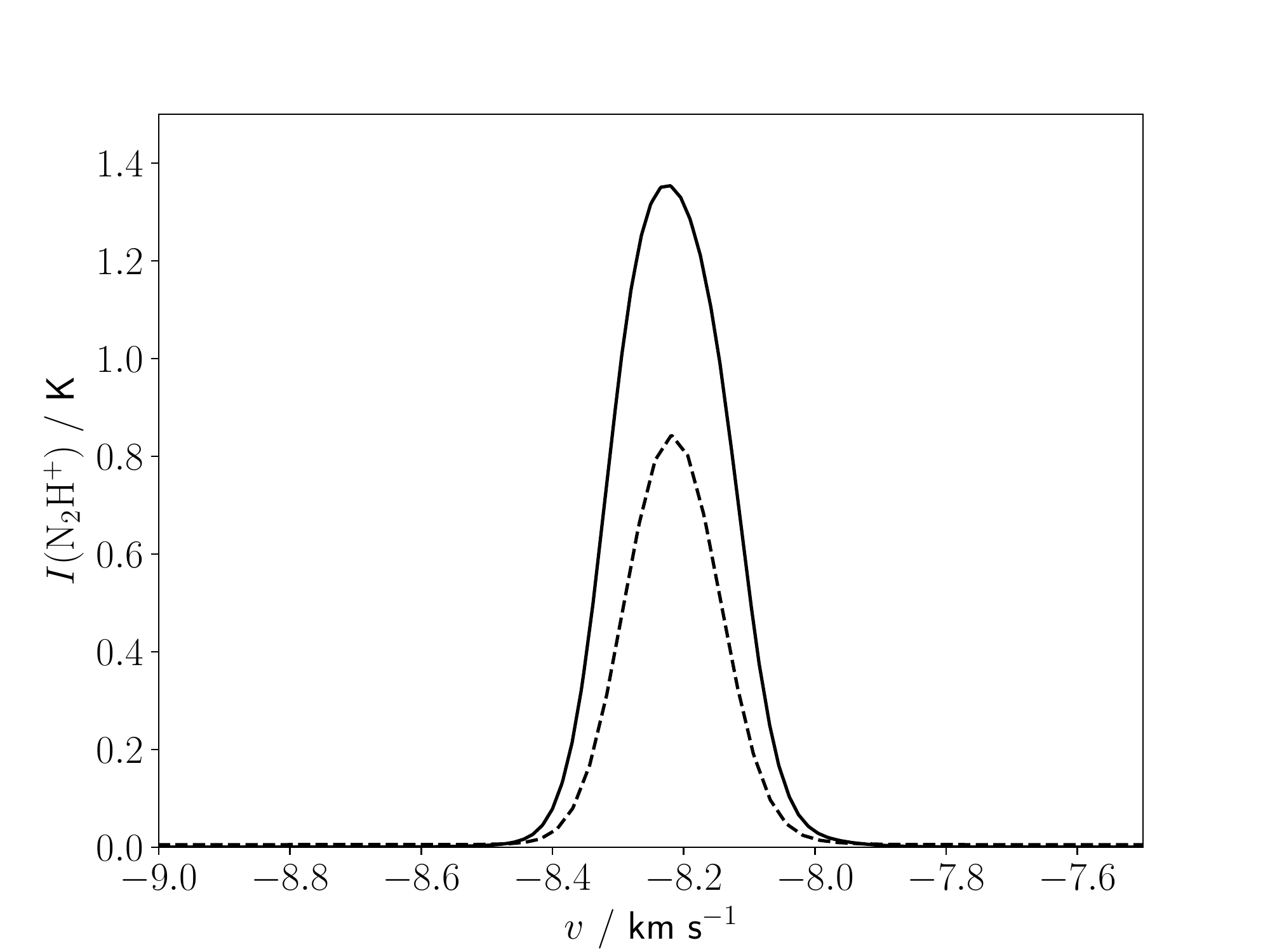}\quad
  \caption{{As Figure \ref{fig:hfs}, but zoomed-in on the central HCN (left) and isolated N$_2$H$^+$ (right) hyperfine components (dashed lines). The non-hyperfine profiles (solid lines) have been shifted to the same central frequency for comparison.}}
  \label{fig:hfszoom}
\end{figure*}

Figure \ref{fig:hfs} shows the effect of hyperfine structure on the HCN {and N$_2$H$^+$ $J=1-0$ lines} for the ${\cal M} = 3$ model after $0.99 \myr$. {For HCN, the individual hyperfine components have lower peak intensities than the non-hyperfine treatment, and display less pronounced self-absorption features, as the column density (and thus optical depth) of the molecule in each hyperfine state is lower than the sum of all equal-$J$ states. The situation for N$_2$H$^+$ is more complex, and individual hyperfine components can be either weaker or stronger than the non-hyperfine approximation. However, the dynamical information encoded in the line profiles is essentially unchanged in both cases.}

{Figure \ref{fig:hfszoom} compares the non-hyperfine profiles to the central HCN component and the isolated N$_2$H$^+$ component (commonly used to trace dense gas dynamics; \citealt{hacar2018,barnes2021}). Line shapes, in particular the line widths (and therefore velocity dispersions) and the velocity of the HCN self-absorption feature, are almost identical. Similar conclusions regarding the dynamics of the filament would be reached regardless of which hyperfine component were considered, particularly for the N$_2$H$^+$ line due to its optically-thin nature. As we do not base any of our conclusions on the absolute intensities of either line but their relative variations, we consider it acceptable to neglect the effects of hyperfine structure.}

\bsp	
\label{lastpage}
\end{document}